\documentclass[%
 reprint,
superscriptaddress,
nofootinbib,
amsmath,amssymb,
aps,floatfix,
pra,
]{revtex4-2}

\usepackage{graphicx}
\usepackage{dcolumn}
\usepackage{bm}
\usepackage{cancel}
\usepackage{hyperref}

\usepackage[dvipsnames]{xcolor}
\usepackage[normalem]{ulem}

\newcommand{\ket}[1]{\left\vert#1\right\rangle}

\newcommand{\ketbra}[2]{| #1\rangle \langle #2|}
\begin{document}

\preprint{APS/123-QED}

\title{
Improved noise resistance analysis for path-entangled subspace quantum key distribution 
}

\author{Osk\'{a}r Adam V\'{a}lent}
 \affiliation{Institute of Computer Science, Masaryk University, 602 00 Brno, Czech Republic}%
\author{Matej Pivoluska}
 \affiliation{Institute of Computer Science, Masaryk University, 602 00 Brno, Czech Republic}%
 \affiliation{Institute of Physics, Slovak Academy of Sciences, 845 11 Bratislava, Slovakia}%

\date{\today}
\begin{abstract}
We revisit the practical implementation of high-dimensional quantum key distribution protocol using path entanglement reported in [Phys.Rev.Lett.~127,~110505,~2021]. 
Here we refine the noise robustness analysis performed there by simulating the effect of different types of channel noise on the noiseless experimental data. 
Further, we extend the analysis to two kinds of asymmetric scenarios. 
In the first scenario the entanglement source is placed in Alice's lab and thus Only one noisy channel has to be considered. 
In the second scenario, Bob's detectors are placed at different physical locations and therefore such scenario implements a simple communication network.
Our analysis can also be extended to other experimental platforms and therefore provides a novel framework for fair noise robustness comparison of high-dimensional quantum key distribution protocols.

\end{abstract}
\maketitle

\section{Introduction}
Secure communication is becoming ever more important in today's world.
Quantum key distribution (QKD) is a quantum communication protocol, which allows two users Alice and Bob obtain two identical copies of a bit-string, called a secret key \cite{BENNETT20147,PhysRevLett.67.661}. 
Secret key can in turn be used as a resource for various information-theoretically secure cryptographic protocols in order to secure users' communications (see e.g.~\cite{HAC}). 
Secret key produced by QKD is guaranteed by the laws of quantum physics to be completely unpredictable to any eavesdropper, which sets it apart from the current classical-information solutions based on computational assumptions.

While QKD is arguably the most mature application of quantum information, its real world deployment is still hindered by relatively low secure key rates and short distribution distances.
It has been reported that using photon pairs entangled in high-dimensional degrees of freedom can solve both of these challenges. This is because high-dimensional entanglement can be used to distribute more than a single bit of key per photon \cite{bechmann-pasquinucci2000,PhysRevA.82.030301,Etcheverry2013,Mirhosseini_2015,PhysRevA.96.022317,islam2017provably,Sit:17,Ding2017,PhysRevApplied.11.064058,Islam_2019,PhysRevApplied.14.014051,Zhong_2015} and additionally, high-dimensional entanglement is more robust to noise \cite{PhysRevX.9.041042}. 
However, using high-dimensional entanglement for QKD in practice is not entirely straightforward, due to various limitations of experimental setups.
Especially, while the data tolerates a large amounts of noise before entanglement can no longer be certified, this surviving entanglement is tied to smaller subspaces \cite{PhysRevX.9.041042}. 
Due to this fact a subspace QKD protocol was introduced in \cite{doda2020quantum}. 
In this protocol the key rate is evaluated in several non-overlapping subspaces of size $k$ of a larger $d$-dimensional Hilbert space. 
This idea enables the harnessing of the entanglement present in the subspaces even in very noisy scenarios, but decreases the maximum achievable key rate per photon to $\log_2(k)$. 
The optimal achievable key rate is therefore an intricate function of total dimension $d$, chosen subspace size $k$ and the noise present in the communication channel.
In \cite{PhysRevLett.127.110505} an experimental demonstration of this protocol was performed with the use of path-entangled photon pairs. 
The goal was to study noise robustness of various subspace QKD protocol parameter settings --- total dimension $d\in\{2,4,8\}$ and key subspace dimension $k\in\{2,4\}$.
The comparison of key rates was performed in a symmetric case with the entanglement source placed between the communicating parties and isotropic noise in the channel (see section \ref{sec:noiseSimulaiton} for details).
In this paper we significantly improve the noise robustness analysis of the experimental setup introduced in \cite{PhysRevLett.127.110505}. 
This is achieved by simulating various types of physical noise in the symmetric scenario (see section \ref{sec:symmetric}), as well as extending the analysis to two interesting asymmetric scenarios. 
In the first one entanglement source placed in Alice's lab (see section \ref{sec:asymmetric}) and in the second one Bob's detection device is physically distributed to multiple locations thus creating a simple QKD network (see section \ref{sec:network}).
We show that in almost all scenarios, the high-dimensional entanglement-based subspace QKD protocol outperforms the traditional qubit protocols.
Our analysis can also be extended to other experimental platforms and therefore provides a novel framework for fair noise robustness comparison of high-dimensional quantum key distribution protocols.

\section{Simulating noise in path-entangled subspace HD-QKD protocol}\label{sec:noiseSimulaiton}

Before describing our noise simulation method, let us briefly introduce the experimental setup used in \cite{PhysRevLett.127.110505}.
{The entangled photon pairs are} created in Type-II spontaneous parametric down-conversion (SPDC) \cite{ou2007multi,schneeloch2019introduction} in which the pump beam is coherently split into $d$ paths with the use of beam-displacers, before hitting the BBO crystal. 
There some of the pump photons are down-converted to an entangled pair of photons. 
The down-converted photon pair is therefore traveling in a superposition of one of $d$ possible paths and it can be described as a state $\ket{\psi_d} = \frac{1}{\sqrt{d}} \sum_{i=0}^{d-1}\ket{ii}$ in a $d\times d$ dimensional Hilbert space. 
The two entangled photons are subsequently separated using a polarizing beam-splitter and send to two different parties, called Alice and Bob, who perform multi-outcome projective measurements on their respective photons. 
The subspace QKD protocol introduced in \cite{doda2020quantum} requires them to switch between the computational basis projective measurement and subspace Fourier-transform basis projective measurements, which are defined, for each $d$ and its divisor $k$ by the orthonormal basis 
\begin{equation}
  \left\{\ket{\bar{i}_j}\bigg\vert \, i\in\{0,1,\dots,k-1\},j\in\{0,1,\dots,\frac{d}{k}-1\}\right\},
\end{equation}
where $\ket{\bar{i}_j} = \frac{1}{\sqrt{k}} \sum_{l=0}^{k-1} \omega_k^{li}\ket{jk+l}$ and $\omega_k = e^\frac{2\pi i}{k}$ is the primitive $k^\mathrm{th}$ root of unity.

Switching between measurement settings can be realized by changing the angles of specific half-wave plates (HWPs) in the paths of the photons.
Importantly, since there are $d$ paths, each party requires $d$ detectors to perform the measurements.
Additionally, in order to study noise resistance of the protocol, each detector coupler in the experiment is equipped with an adjustable intensity light
emitting diode (LED) which is used to introduce noise to each detector. 
Note that this high-level description of the experimental setup is sufficient for the work presented in this paper and we refer the interested reader to \cite{PhysRevLett.127.110505} for technical details.

The basic data unit produced in the experiment takes a form of two $d\times d$ coincidence matrices $C_\mathrm{Comp}$ and $C_\mathrm{Four}$, which in $i$-th row and $j$-th column contain the number of simultaneous (within the same coincidence window set to $\tau = 5\times 10^{-9}$~s) detector clicks produced by $i$-th Alice's and $j$-th Bob's detector during the run of the experiment, when simultaneously measuring in the computational basis or the subspace Fourier basis respectively.
In the ideal noiseless experiment both of these matrices are proportional to identity.
The role of the noise LEDs is to produce additional single click events in each detector with a precisely controllable rate $S$, which sometimes result in accidental coincidences. These extra coincidences manifest as noise in the data matrices $C_\mathrm{Comp}$ and $C_\mathrm{Four}$ and they are distributed uniformly, i.e. any pair of detectors has equal probability $\frac{1}{d^2}$ to produce an accidental coincidence.

In \cite{PhysRevLett.127.110505}, this setup is used to simulate an experimental scenario, in which an isotropic state 
\begin{align}
\Psi_d = (1-p)\ketbra{\psi_d}{\psi_d} + p\frac{I_{d^2}}{d_2}
\end{align}
is distributed to the communicating parties. 
Isotropic states are a useful tool to study noise robustness of various protocols and entanglement properties. They can be seen as a classical mixture of the maximally entangled state $\ketbra{\psi_d}{\psi_d}$, with the fully mixed state $\frac{I_{d^2}}{d_2}$ in $d\times d$-dimensional Hilbert space and are parametrized by the probability $p\in [0,1]$ of the state being fully mixed.
In \cite{PhysRevLett.127.110505}, five levels of noise were experimentally compared, corresponding to   $p\in\{0,0.025,0.075,0.15,0.3\}$.

In this paper we take a different approach to studying the noise resistance of the subspace QKD protocol implemented in \cite{PhysRevLett.127.110505}. 
The main idea is that the effect of the LED noise sources on the coincidence matrices produced in the experiment is  predictable, and thus, with access to the noiseless data generated in the experiment, i.e.~matrices $C_\mathrm{Comp}$ and $C_\mathrm{Four}$ with $p=0$ setting, various  interesting noise scenarios can be studied.
The advantage of this approach compared to the more traditional full statistical modeling of quantum photonic setups (see e.g. \cite{PhysRevA.76.012307,PhysRevA.104.022406}) is that some noise sources originating from device imperfections, such as dark counts of detectors or polarization error induced by imperfect optical components, are partially factored in the noiseless data and only channel noise needs to be simulated.

The basic tool of the channel noise simulation is a formula that calculates the rate of accidental coincidences resulting from the single click events introduced to the detectors.
In order to encompass asymmetric scenarios studied in this paper, we assume that single click events are introduced into each detector in Alice's lab with rate $S_A$ and into Bob's lab with with rate $S_B$. 
The first simplification is that we assume both Alice's and Bob's labs obtain $dS_A$ and $dS_B$ extra single clicks over all $d$ detectors.
In other words, we neglect possible multi-click events occurring in a single lab. 
Such a simplification is justifiable, because with the coincidence window $\tau = 5\times 10^{-9}$ and Hilbert space dimensions up to $8$, multi-click rates are several orders of magnitude lower than the critical accidental coincidence rates (i.e.~coincidence rates that cause the key rate to vanish).
The accidental coincidence rate $C(S_A,S_B,d)$ with $d$ detectors on each side and extra single count rates $S_A$ and $S_B$ then can be calculated as 
\begin{align}
    C(S_A,S_B,d) &\!=\!dS_A(1\!-\!e^{-\tau dS_B}) + dS_B(1\!-\!e^{-\tau dS_A}),
\end{align}
where $1-e^{-\tau dS_B}$ and $1-e^{-\tau dS_A}$ are the probabilities that Bob and Alice respectively observe a single count in time interval $\tau$. Since we are working in regimes with $\tau \ll S_A$ and $\tau \ll S_B$, both  $(1-e^{-\tau dS_A})$ and $(1-e^{-\tau dS_B})$ can be expanded to their first order as 
$1 - e^{-\tau dS_A} \approx 1 - (1 - \tau dS_A) =  \tau dS_A$ and $1 - e^{- \tau dS_B} \approx 1 - (1 - \tau dS_B) = \tau dS_B$. Thus the extra coincidence count rate can be estimated as
\begin{align}\label{eq:extra counts}
        C(S_A,S_B,d) & \approx 2\tau d^2S_AS_B.
\end{align}

In the next section we will use eq.~\eqref{eq:extra counts} in conjunction with noiseless coincidence count matrices $C_\mathrm{Comp}$ and $C_\mathrm{Four}$ obtained from the experiment performed in \cite{PhysRevLett.127.110505} (see Appendix \ref{app:matrices}) in order to study various types of physically inspired noise in different practical subspace QKD setups.
In most cases this is done by adding a randomly generated noise matrix $C_\mathrm{Noise}$ to the coincidence matrices $C_\mathrm{Comp}$ and $C_\mathrm{Four}$. 
Since the added accidental coincidences are uniformly distributed for each noise level, different $C_\mathrm{Noise}$ are generated by uniformly distributing $25\times C(S_A,S_B,d)$ accidental counts into a $d\times d$ matrix.
Note that multiplication by $25$ to obtain the amount of accidental coincidences is performed because the duration of all experiments performed in \cite{PhysRevLett.127.110505} was $25$ seconds.
Types of noise studied in this paper differ mainly in how many additional single counts $S_A$ and $S_B$ they produce, since this quantity depends on the scenario and the total number of deterctors in each lab.
Our results therefore can be interpreted as a more elaborate way to compare noise robustness of instances of subspace QKD protocol with different total dimensionality $d$ and subspace size $k$.

\section{Results}
Our results are split into three parts grouped by the practical scenario employed. In the first part we study a symmetric scenario with the entanglement source between the communicating parties.
Second part is devoted to an asymmetric scenario with the entanglement source in Alice's lab.
In the last part we study a novel potential application of path-entanglement --- an implementation of the subspace QKD protocol, in which Bob's detectors are split into three separate labs. Such a setup implements a simple QKD network scenario.

\begin{figure}[ht!]
    \vspace{-0.3cm}%
\centering
 {
        \includegraphics[width=0.45\textwidth]{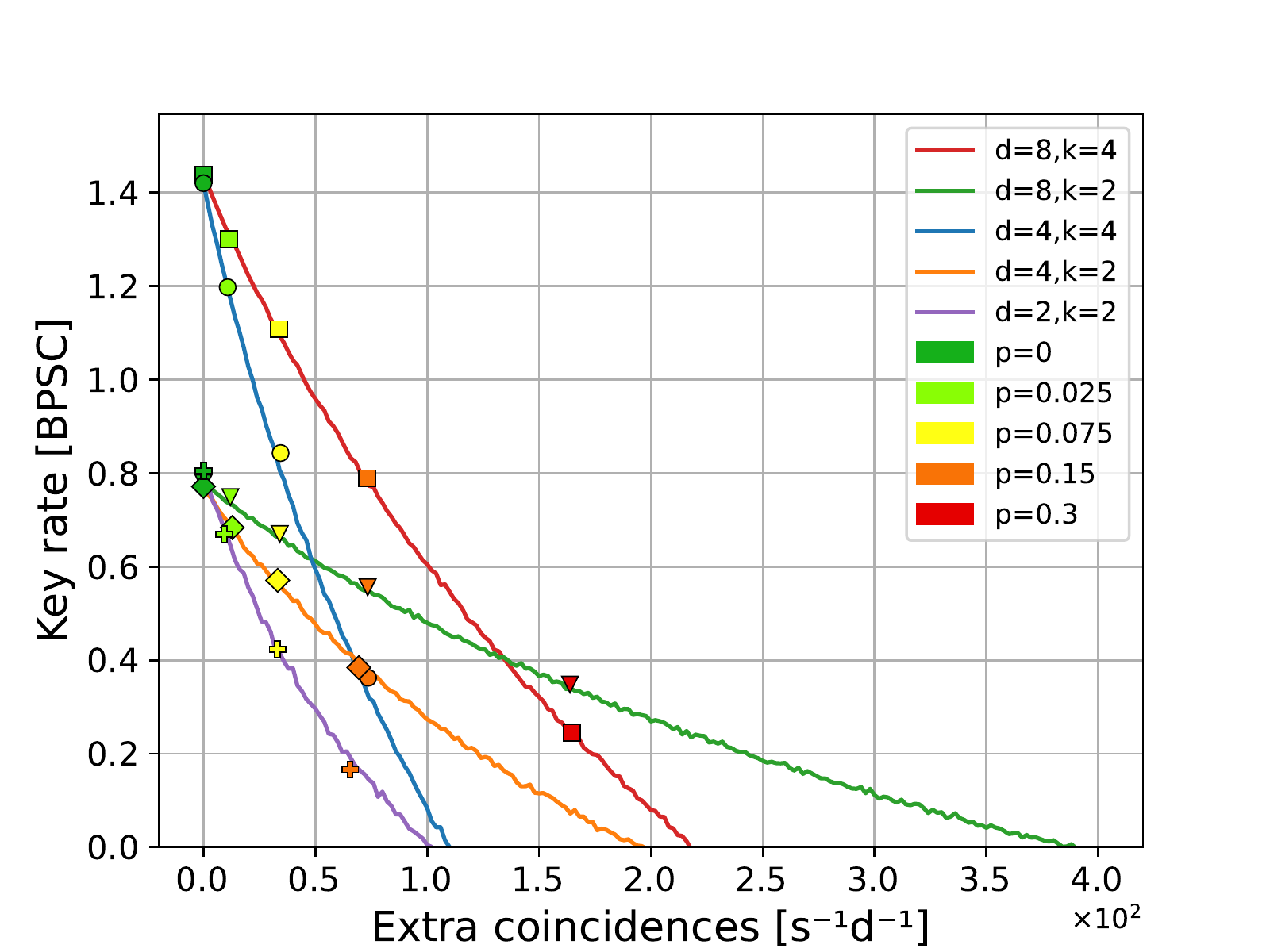}
        
        (a) Source in the middle -- Isotropic noise.
    }
    {
        \includegraphics[width=0.45\textwidth]{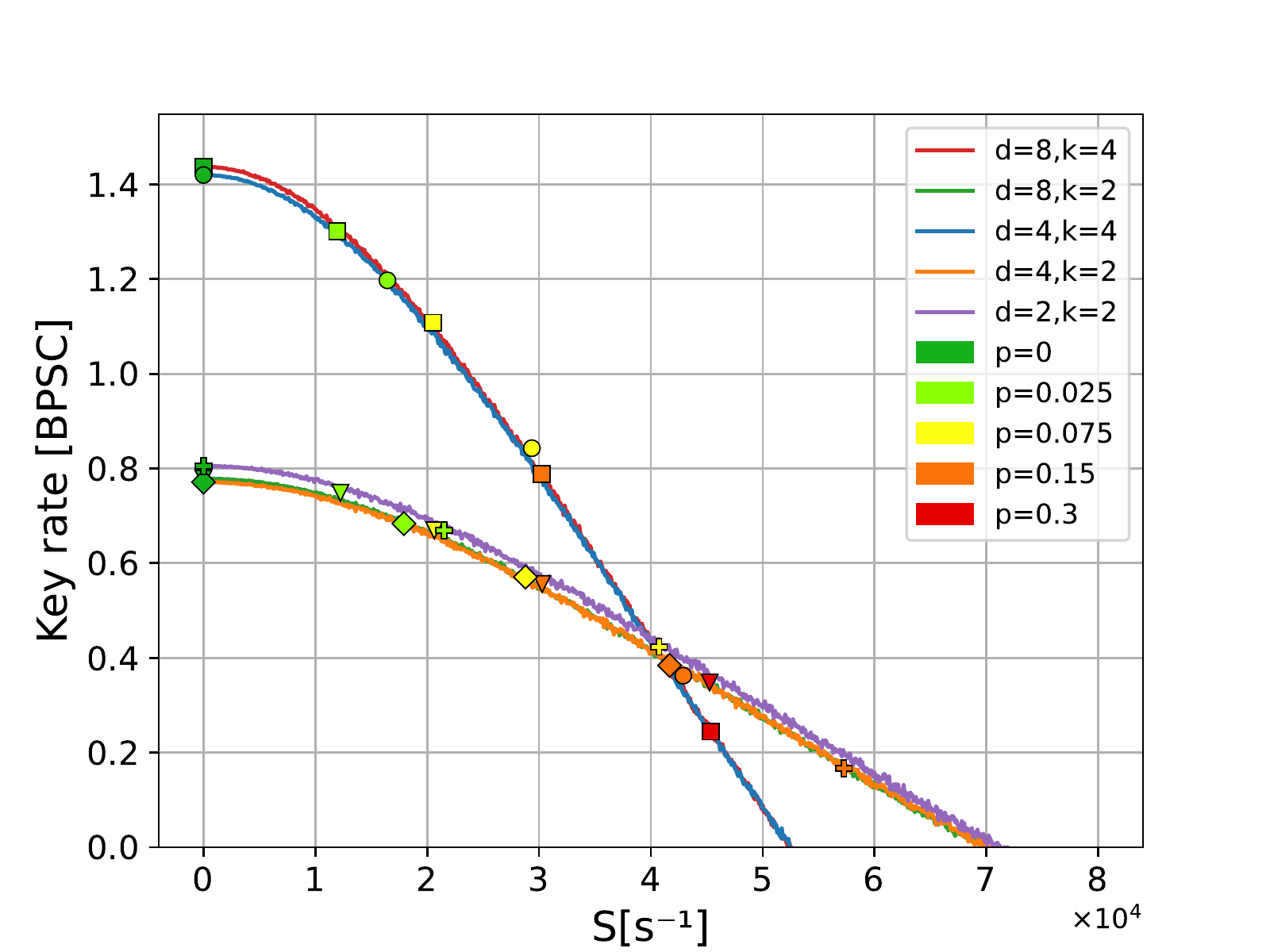}
        
        (b) Source in the middle -- Detector noise.

    }
   {
        \includegraphics[width=0.45\textwidth]{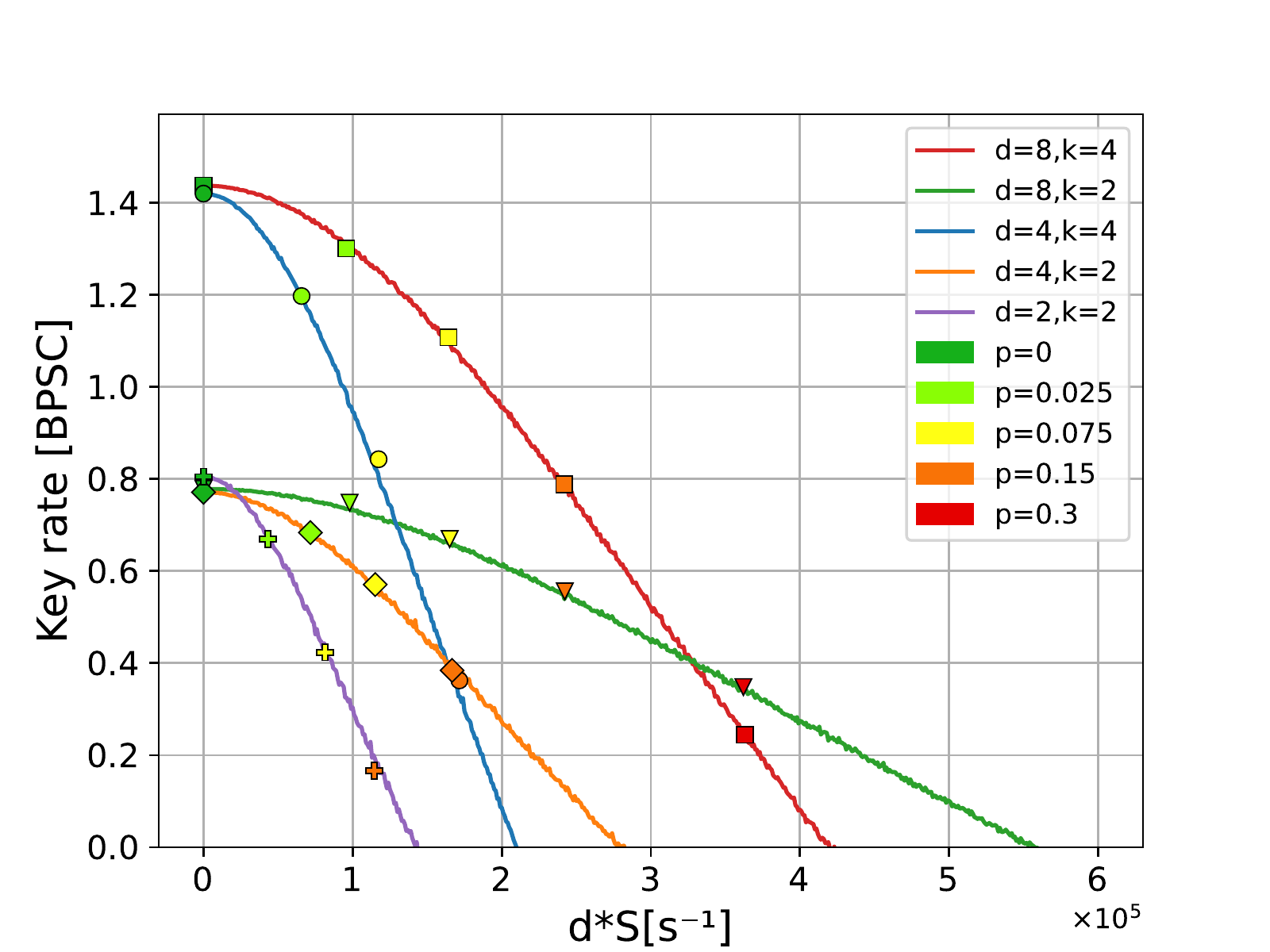}
        
        (c) Source in the middle -- Channel noise.
      
    }
    \caption{Simulated key rates in case of symmetric scenario In (a) we compare key rates of subspace QKD protocols with different settings of $d$ and $k$ based on the isotropic noise parameter $p$. In (b)  we use the single count rate of each detector as the measure of noise. In (c)  we use total single count rate per party $dS$ as the measure of noise. In order to compare our results with the experimental results produced in \cite{PhysRevLett.127.110505}, we plot the obtained experimental points as well.}
    \label{fig::globfig}
\end{figure}

\subsection{Entanglement source in the middle} \label{sec:symmetric}
In this subsection we reexamine the scenario studied in \cite{PhysRevLett.127.110505}, in which the entanglement source is placed between Alice's and Bob's labs and therefore $S_A = S_B = S$.
There are three physically inspired types of noise we study here.
First, we derive the results for isotropic noise for a full range of noise parameter $p$. 
Interestingly, as pointed out in \cite{PhysRevLett.127.110505}, isotropic noise parameter $p$ is proportional to an accidental coincidence rate $C(S,S,d)/d = 2\tau d S^2$.
Therefore to compare the key rates at the same isotropic noise level $p$ for instances 
of QKD with different total dimension, individual single rates $S$ decrease with $d$.
This prompts a question whether isotropic noise is a fair way to compare instances of the high-dimensional QKD protocols with different $d$.
To provide a more fair comparison, we propose two additional types of noise inspired by the actual physical implementation.
The first type, we call \emph{detector noise},  is characterized by the rate of single counts $S$. 
This type of noise is consistent with increasing the dark count rate of the detectors.
Second type, we call \emph{channel noise}, is characterized by the total single count rate $dS$ introduced to each lab.
This is physically motivated by outside light entering the labs from the communication channel, which is a relevant scenario for free-space experiments.

In FIG. \ref{fig::globfig} we plot the key rate per subspace coincidence (BPSC) (i.e.~key bits per photon in the asymptotic scenario) for all three noise types.
Additionally, we plot the experimental points from \cite{PhysRevLett.127.110505} to show that the models are accurate.
Interestingly, it is revealed that in presence of detector noise, increasing the total dimensionality $d$ does not provide any advantage in BPSC and only increasing $k$ effectively changes the key rate in low noise regimes. 
On the other hand, in presence of channel noise the advantage of using higher $d$ is even more pronounced than in the case of isotropic noise.
This has an intuitive explanation -- photons entering the measurement setup from the channel are split equally between all the detectors, therefore increasing the number of detectors in the setup results in smaller individual single count rates in each detector.
In FIG. \ref{fig::globfig}(c) this can be seen from the fact that the experimental points of the same color, which were obtained in experimental runs with different $d$ and $k$, but the same $p$, are not vertically aligned anymore.

\subsection{Entanglement source in Alice's lab}\label{sec:asymmetric}
\begin{figure}[h!]
    \vspace{-0.3cm}%
    \begin{center}
        
    {
        \includegraphics[width=0.45\textwidth]{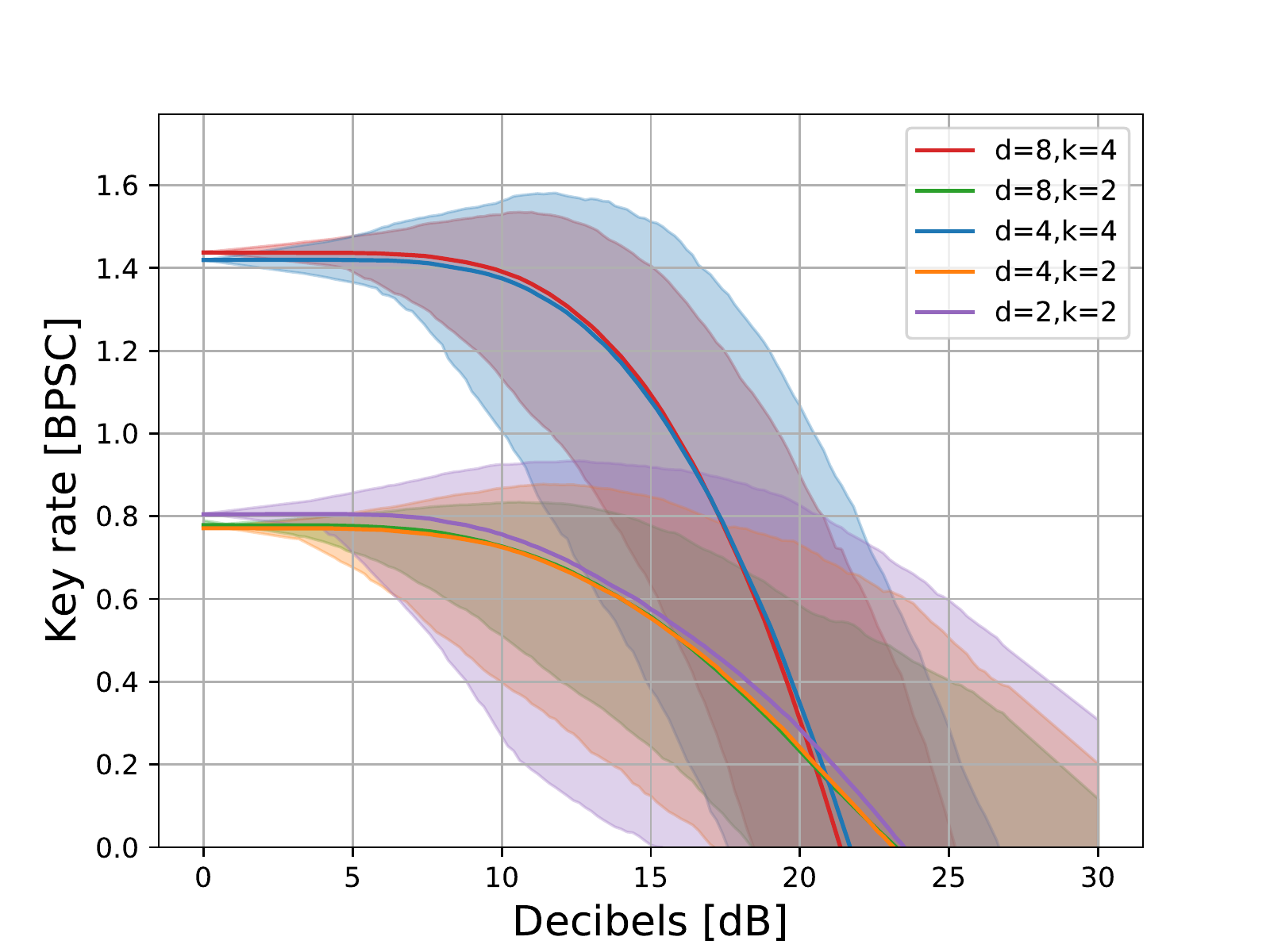}
        
        Source in Alice's lab with $dS_b = 0$
    }
        
    {
        \includegraphics[width=0.45\textwidth]{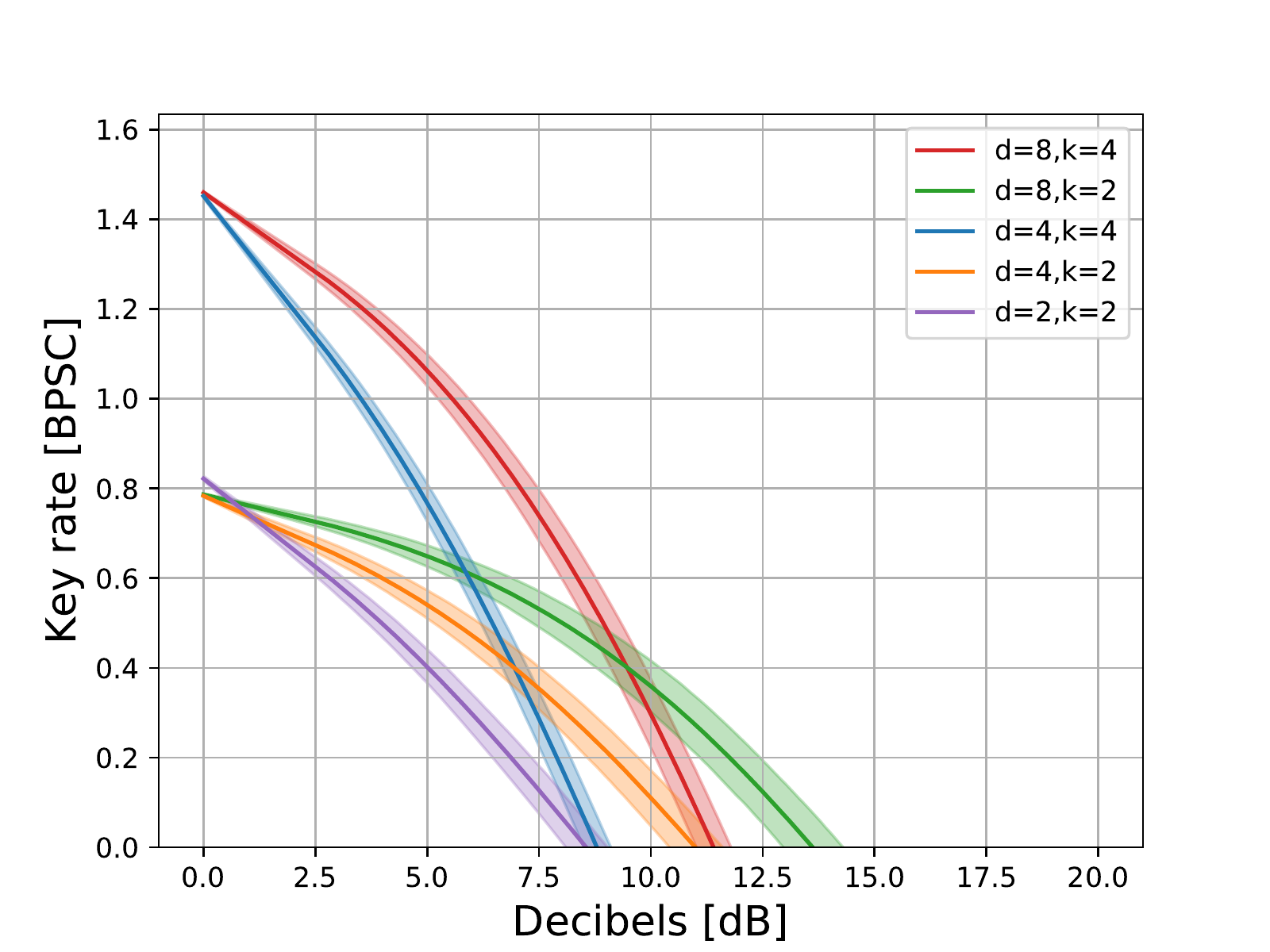}
       
       Source in Alice's lab with $dS_b = 160k$
    }
    
      {
        \includegraphics[width=0.45\textwidth]{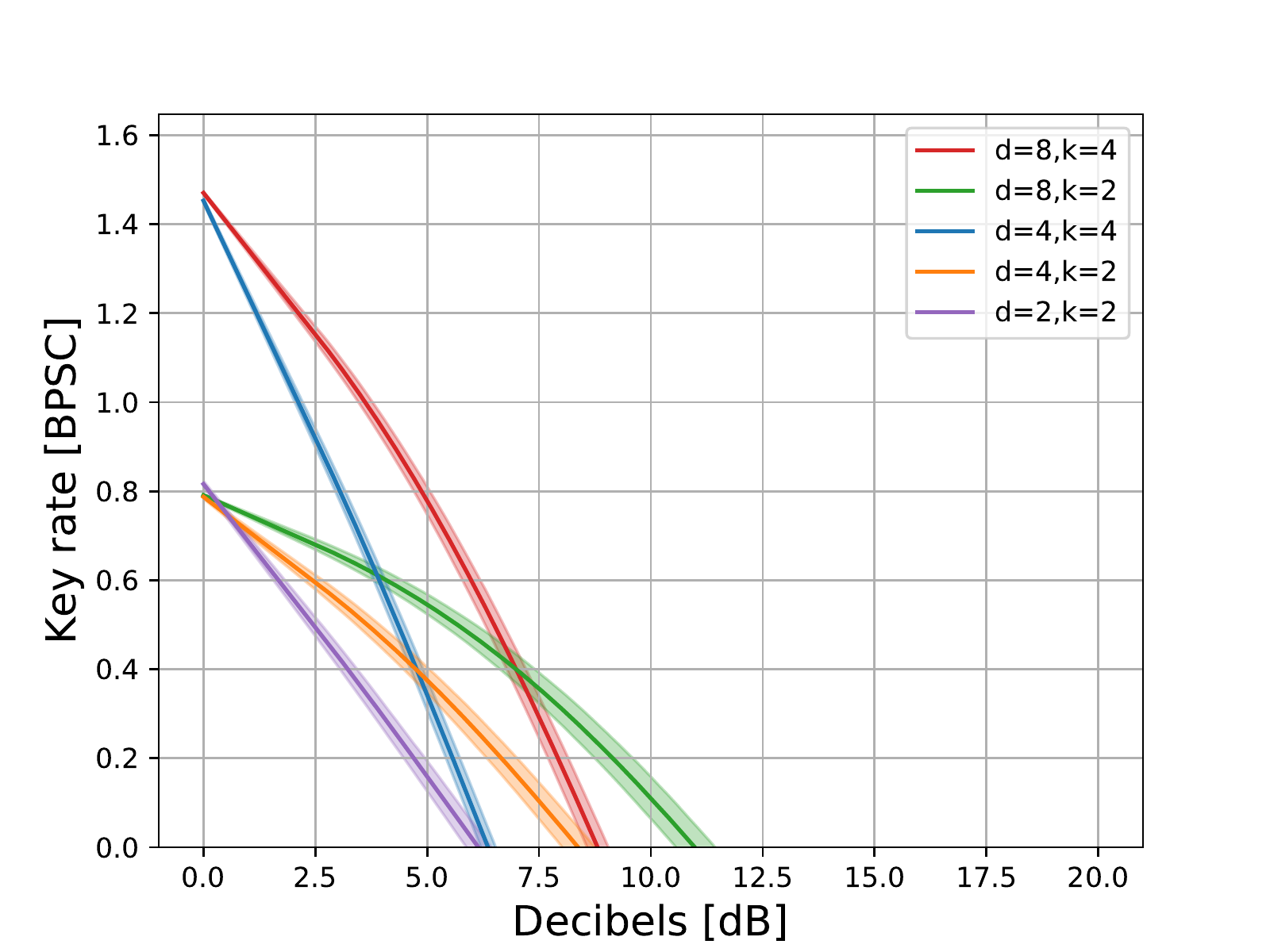}
        
        Source in Alice's lab with $dS_b = 320k$
    }
    \end{center}
    \caption{Graphs for the scenario with the entanglement source placed in Alice's lab. Part (a) shows the effect of channel noise without outside noise contribution from the channel, i.e.~there are no photons entering Bob's lab. In (b) and (c) the channel noise parameter $dS_B$ was set to $160000$ and $320000$ respectively. The line in the middle represents an average BPSC obtained over $100$ simulations, while top and bottom line are the best/worst key rate observed respectively.}
    \label{fig:asymetric}
\end{figure}

In this subsection we study an asymmetric scenario, in which the entanglement source is placed in Alice's lab. 
In this scenario the amount of single counts in Alice's and Bob's lab is not symmetric anymore, because their physical sources differ. 
While the source of singles for Bob is the detector noise and channel noise, just like in the symmetric case discussed above, we assume channel noise does not contribute to Alice's single counts anymore, since she has a full control of the channel between the source and her detectors. 
Instead, we take into account channel loss --- Bob's part of the entanglement pair can get lost in the channel. 
This type of physical noise has two contributions to the coincidence count matrices.
First, the number of counts originating from the entanglement source decreases. We model this by randomly erasing counts from the noiseless matrices $C_\mathrm{Comp}$ and $C_\mathrm{Four}$. 
At the same time, for each source photon lost in the channel, Alice registers a single count, which contributes to her single count rate $S_A$.
In FIG. \ref{fig:asymetric} we plot the obtainable key rate in BPSC for different settings of $d$ and $k$ in a subspace QKD protocol, with the channel loss in $dB$ as the noise parameter.
Here, we set the detector noise count rate to $0$, as we assume that the intrinsic dark count rate of the detectors used in the experiment is already accounted for in the noiseless matrices $C_\mathrm{Comp}$ and $C_\mathrm{Four}$, but it can in principle also be set to an arbitrary value.
To account for channel noise in Bob's lab the figure contains graphs for three different levels of channel noise $dS_B\in\{0,160000,320000\}$ contributing singles on Bob's side. 
Because erasing clicks from the data matrices increases the variance of the simulated key rates (especially in the case of $dS_B = 0$,
we performed each simulation $100$ times and plot the average value (thick lines in FIG.\ref{fig:asymetric}), as well as minimum and maximum value observed, thus giving error bounds on the key rate for each noise parameter value.

In this case the results show that channel loss behaves similarly to the detector noise -- there is no advantage in increasing the local dimension $d$ with the same subspace size $k$. 
However, combining this type of noise 
with the channel noise again suggest that increasing $d$ while keeping $k$ fixed increases the key rates for all studied noise levels. 

\subsection{QKD network via path entanglement}\label{sec:network}
In this subsection we introduce a novel scenario enabled by the path entanglement implementation of the subspace QKD protocol and simulate the effect of channel noise on obtainable key rates.
Since paths each photon in the entangled pair are spatially separated, it is possible to divide  Bob's detectors into subsets and place them in different physical locations. 
In such a layout it would still be possible to perform the measurements required for certain variants of the subspace QKD protocol, but each subspace could be controlled by a different entity. 
In practice this would implement a simple QKD network with a star architecture.

\begin{figure}[ht!]
    \vspace{-0.3cm}%
    \begin{center}
        
    {
        \includegraphics[width=0.45\textwidth]{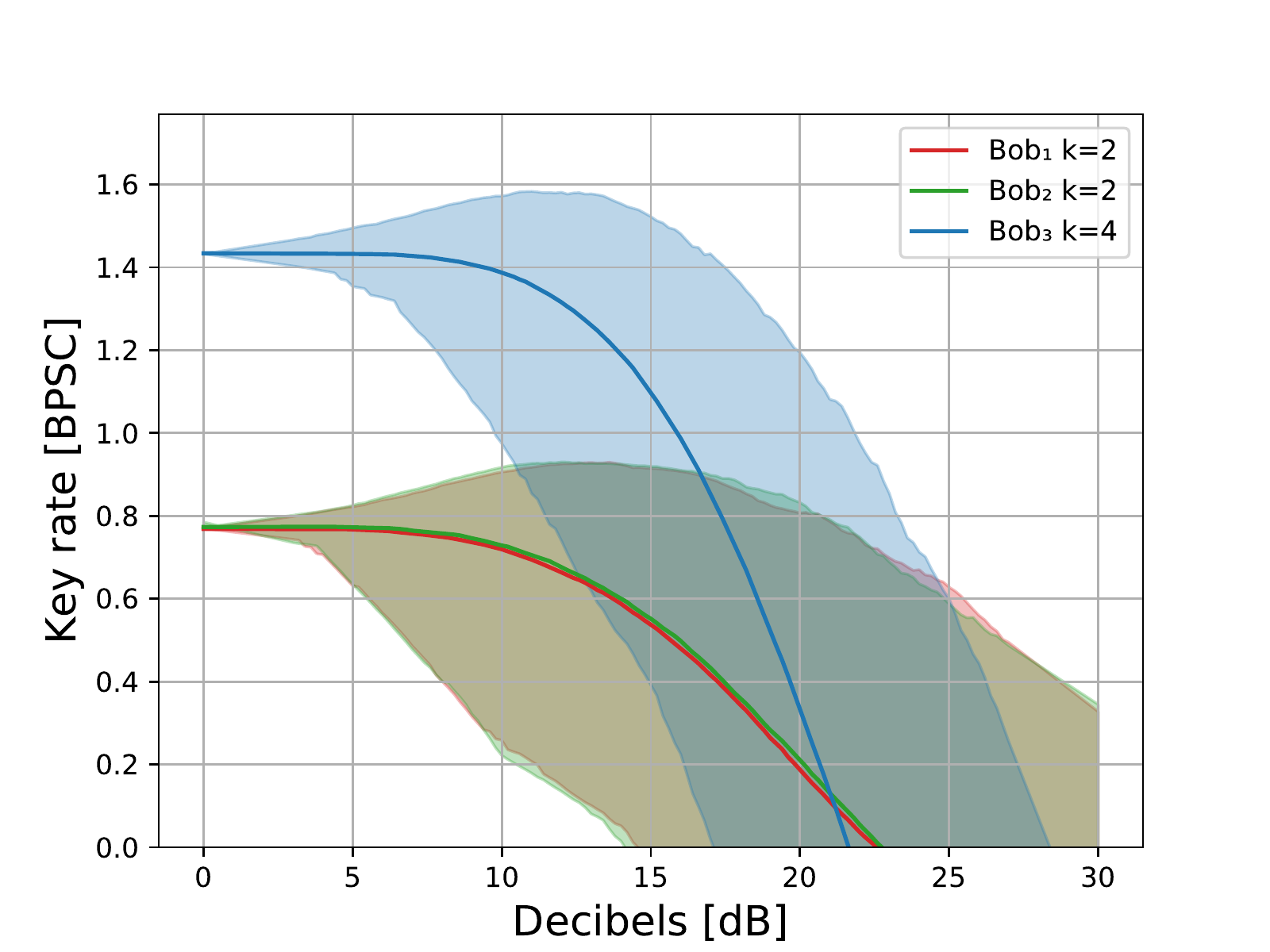} 
        
        Star network with $F = 0$.

    }

    {
        \includegraphics[width=0.45\textwidth]{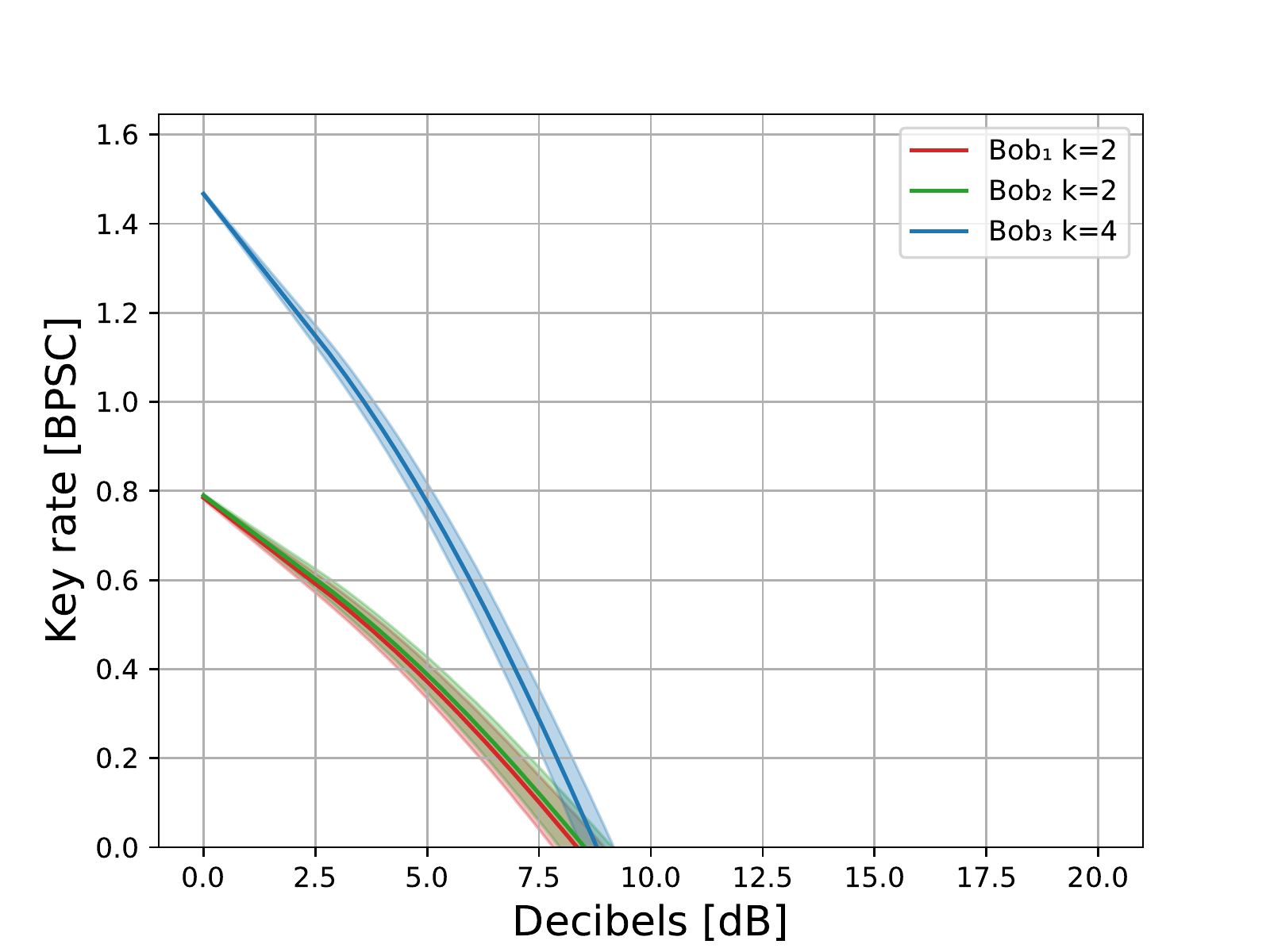}
        
        Star network with $F = 160k$.

    }

    {
        \includegraphics[width=0.45\textwidth]{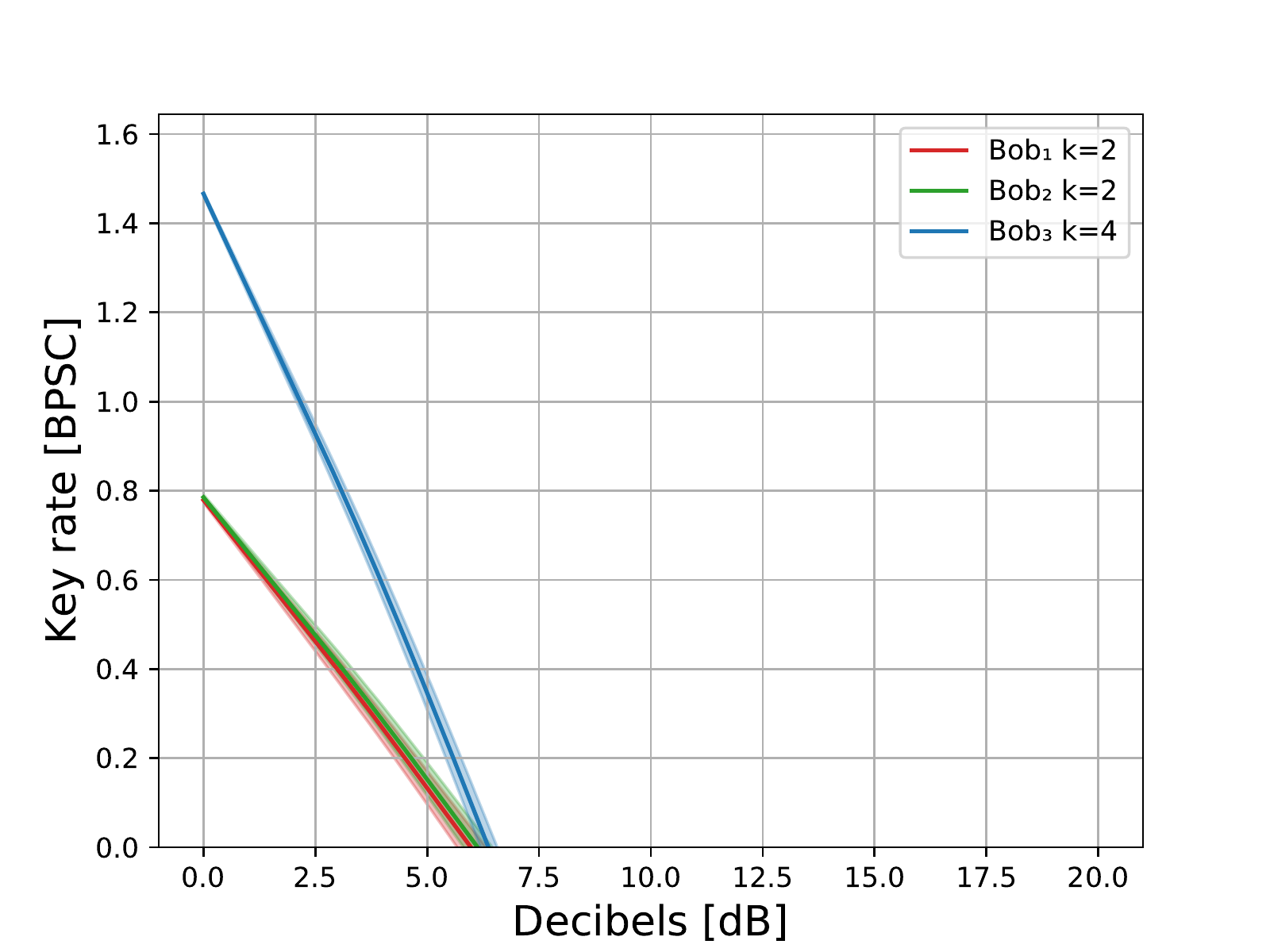}
        
        Star network with $F = 320k$.
 
    }
    
    \end{center}
    
    \caption{QKD network key rates. Bob's detectors are split into three labs. Parameter $F$ characterizes the outside channel conditions and represent the rate of the number photons entering each lab in one second. The line in the middle represents an average BPSC obtained over $100$ simulations, while top and bottom line are the best/worst key rate observed respectively. }
    \label{fig:network}
\end{figure}

Similar networks were implemented before \cite{Wengerowsky2018,Joshi2020}, using photons entangled simultaneously in frequency and polarization degrees of freedom.
Here we exemplify the scenario by considering an entanglement source with $8$ paths (i.e.~$d=8$) and splitting Bob's detectors into three physically distinct locations. 
The first two locations (Bob$_1$ and Bob$_2$) control $2$ detectors each, while the remaining location ( Bob$_3$) controls $4$ detectors.
Since the combination of measurements required for this setup was not performed in the experiment, we construct the coincidence matrices $C_\mathrm{Comp}$ and $C_\mathrm{Four}$ from coincidence matrices obtained in $d=8, k=2$ and $d=8, k=4$ cases, by taking the blocks from the corresponding subspaces (see Appendix \ref{app:matrices}).
As for the noise, we assume the entanglement source is placed in Alice's lab. 
Like in the asymmetric case with two parties, the only contribution to $S_A$ is from the channel loss. 
Difference to the two party scenario is that the same number of extra photons from the channel needs to be added to each physical location with the detectors, which is consistent with outside light entering the setup in a free-space experiment.
This in particular means that the single count rate for Bob$_3$'s detectors $S_{B_3} = \frac{1}{2}S_{B_1} = \frac{1}{2}S_{B_2}$, where $S_{B_1}$ and $S_{B_2}$ are single count rates of Bob$_1$ and Bob$_2$ respectively. 
This ensures that $4S_{B_3} = 2S_{B_1} = 2S_{B_2}$, i.e.~the number of photons entering each lab is the same. 
Because of this asymmetry, we use the coincidence count formula of Eq.\eqref{eq:extra counts} separately for each subspace, leading to the extra coincidence formula for each Bob$_i$ of the form:
\begin{align}
    C_{AB_i}(d_i) = 2\times25\times \tau \times \frac{dS_a}{8/d_i}\times F, 
\end{align}
where $F$ is the intensity of the channel noise in single counts per second, $d_i$ is the total local dimension of Bob$_i$ and $dS_A$ is Alice's single count rate resulting from the channel loss.
Results of this simulation are depicted in FIG. \ref{fig:network}. 
Again,
we performed each simulation $100$ times and plot the average value (thick lines in FIG.\ref{fig:network}), as well as minimum and maximum value observed, thus giving error bounds on the key rate for each noise parameter value.
As expected, the results are similar to the results in the asymmetric scenario, as the key rates for Bob$_3$ are basically identical to the $d=4,k=4$ case, while the key rates for Bob$_1$ and Bob$_2$ are identical to the $d=2,k=2$ case. Note, however that in principle Bob$_3$'s BPSC can be made identical to Bob$_1$ and Bob$_2$, since he can freely switch to the $d=4,k=2$ case as well.

\section{Discussion}
In this paper we have revisited the implementation of the high-dimensional subspace QKD protocol in the path degree 
of freedom presented in \cite{PhysRevLett.127.110505}.
We have extended the noise resistance analysis of the implementation to include various noise types 
that were not considered before --- detector noise, channel noise and channel loss. 
In addition, we present a novel network scenario enabled by the path-entanglement and 
analyze its noise resistance as well.
In broader context, we present a more complex way to compare the noise resistance of QKD protocols
implemented with high-dimensional degrees of freedom that goes well beyond considering the 
visibility of the state as the figure of noise. Additionally, the technique of taking the experimental data from 
an (almost-noiseless) tabletop experiment and simulating the effects of different types of 
noise has advantages compared to the full statistical analysis. 
In particular, the noise that originates 
in the optical elements and are independent of the channel do not have to be simulated as they are already accounted for in the data. 

\section{Acknowledgements}
O.~A.~V. and M.~P. acknowledge funding and support from GAMU project MUNI/G/1596/2019. M.P. further acknowledges VEGA Project No. 2/0136/19. We would like to thank Xiaomin Hu, Marcus Huber and Lukas Bulla for valuable discussions.

\section{References}

\clearpage
\onecolumngrid
\appendix
\section{Noiseless matrices}\label{app:matrices}
In this appendix we list the experimental coincidence matrices obtained in the experiment reported in \cite{PhysRevLett.127.110505}, which were used as a basis of our simulations. All of them were collected in experimental runs with duration of $25$~s.

\begin{table}[h!]
    \centering
    \begin{tabular}{|c|c|c|c|c|c|c|c|c|} 
        \hline
        Comp. & $\ket{0}$ & $\ket{1}$ & $\ket{2}$ & $\ket{3}$ & 
           $\ket{4}$ & $\ket{5}$ & $\ket{6}$ & $\ket{7}$ \\ [0.5ex] 
        \hline
        $\ket{0}$ & 10372 & 42 & 51 & 31 & 8 & 2 & 3 & 12 \\ 
        \hline
        $\ket{1}$ & 36 & 9641 & 37 & 41 & 6 & 7 & 15 & 2 \\
        \hline
        $\ket{2}$ & 41 & 19 & 10821 & 45 & 6 & 13 & 10 & 6 \\
        \hline
        $\ket{3}$ & 34 & 65 & 41 & 10311 & 7 & 5 & 14 & 9 \\
        \hline
        $\ket{4}$ & 7 & 8 & 7 & 6 & 9715 & 42 & 53 & 41 \\  
        \hline 
        $\ket{5}$ & 4 & 5 & 13 & 9 & 21 & 9621 & 32 & 69 \\
        \hline
        $\ket{6}$ & 7 & 4 & 5 & 11 & 47 & 23 & 10571 & 37 \\
        \hline
        $\ket{7}$ & 11 & 14 & 8 & 7 & 31 & 71 & 41 & 10003 \\
        \hline
    \end{tabular}
    \caption{Noiseless computational coincidence matrix for $d=8$.}
    \label{tab:my_label}
\end{table}

\begin{table}[h!]
    \centering
    \begin{tabular}{|c|c|c|c|c|c|c|c|c|} 
        \hline
        Four. & $\ket{0}_1$ & $\ket{1}_1$ & $\ket{0}_2$ & $\ket{1}_2$ & 
           $\ket{0}_3$ & $\ket{1}_3$ & $\ket{0}_4$ & $\ket{1}_4$ \\ [0.5ex] 
        \hline
        $\ket{0}_1$ & 9569	& 311 & 70 & 106 & 7 & 6 & 2 & 10 \\ 
        \hline
        $\ket{1}_1$ & 290 & 9768 & 76 & 41 & 6 & 7 & 12 & 5 \\
        \hline
        $\ket{0}_2$ & 42 & 134 & 10315 & 318 & 3 & 12 & 5 & 7 \\
        \hline
        $\ket{1}_2$ & 162 & 54 & 271 & 9518 & 6 & 4 & 4 & 10 \\
        \hline
        $\ket{0}_3$ & 7 & 3 & 10 & 4 & 9218 & 253 & 32 & 78 \\  
        \hline 
        $\ket{1 }_3$ & 13 & 16 & 14 & 8 & 270 & 9401 & 102 & 72 \\
        \hline
        $\ket{0}_4$ & 6 & 2 & 10 & 10 & 67 & 103 & 9625 & 279 \\
        \hline
        $\ket{1}_4$ & 8 & 11 & 7 & 4 & 178 & 35 & 254 & 9577 \\
        \hline
    \end{tabular}
    \caption{ Noiseless subspace Fourier coincidence matrix for $d=8,k=2$.}
    \label{tab:my_label}
\end{table}

\begin{table}[h!]
    \centering
    \begin{tabular}{|c|c|c|c|c|c|c|c|c|} 
        \hline
        Four. & $\ket{0}_1$ & $\ket{1}_1$ & $\ket{2}_1$ & $\ket{3}_1$ & 
           $\ket{0}_2$ & $\ket{1}_2$ & $\ket{2}_2$ & $\ket{3}_2$ \\ [0.5ex] 
        \hline
        $\ket{0}_1$ & 9541 & 305 & 307 & 89 & 10 & 3 & 5 & 7 \\ 
        \hline
        $\ket{1}_1$ & 337 & 9321 & 83 & 271 & 7 & 10 & 9 & 7 \\
        \hline
        $\ket{2}_1$ & 203 & 72 & 9314 & 303 & 4 & 6 & 18 & 9 \\
        \hline
        $\ket{3}_1$ & 104 & 337 & 251 & 9931 & 6 & 6 & 3 & 19 \\
        \hline
        $\ket{0}_2$ & 8 & 12 & 13 & 10 & 9213 & 301 & 250 & 75 \\  
        \hline 
        $\ket{1}_2$ & 4 & 10 & 10 & 13 & 299 & 9437 & 107 & 261 \\
        \hline
        $\ket{2}_2$ & 11 & 5 & 5 & 7 & 311 & 103 & 9321 & 307 \\
        \hline
        $\ket{3}_2$ & 8 & 12 & 10 & 4 & 108 & 307 & 201 & 9632 \\
        \hline
    \end{tabular}
    \caption{Noiseless subspace Fourier coincidence matrix for $d=8,k=4$.}
    \label{tab:my_label}
\end{table}

\begin{table}[h!]
    \centering
    \begin{tabular}{|c|c|c|c|c|} 
        \hline
        Comp. & $\ket{0}_1$ & $\ket{1}_1$ & $\ket{0}_2$ & $\ket{1}_2$ \\ [0.5ex] 
        \hline
        $\ket{0}_1$ & 10562	& 53 & 45 & 21 \\ 
        \hline
        $\ket{1}_1$ & 47 & 9731 & 32 & 43 \\
        \hline
        $\ket{0}_2$ & 57 & 18 & 10714 & 41 \\
        \hline
        $\ket{1}_2$ & 37 & 71 & 35 & 10136 \\
        \hline
    \end{tabular}
    \caption{Noiseless computational coincidence matrix for $d=4$.}
    \label{tab:my_label}
\end{table}

\begin{table}[h!]
    \centering
    \begin{tabular}{|c|c|c|c|c|} 
        \hline
        Four. & $\ket{0}_1$ & $\ket{1}_1$ & $\ket{0}_2$ & $\ket{1}_2$ \\ [0.5ex] 
        \hline
        $\ket{0}_1$ & 9404 & 321 & 68 & 143 \\ 
        \hline
        $\ket{1}_1$ & 291 & 9935 & 99 & 39 \\
        \hline
        $\ket{0}_2$ & 53 & 201 & 10235 & 251 \\
        \hline
        $\ket{1}_2$ & 105 & 43 & 304 & 9523 \\
        \hline
    \end{tabular}
    \caption{Noiseless subspace Fourier coincidence matrix for $d=4,k=2$.}
    \label{tab:my_label}
\end{table}

\begin{table}[h!]
    \centering
    \begin{tabular}{|c|c|c|c|c|} 
        \hline
        Four. & $\ket{0}_1$ & $\ket{1}_1$ & $\ket{2}_1$ & $\ket{3}_1$ \\ [0.5ex] 
        \hline
        $\ket{0}_1$ & 9550 & 321 & 307 & 104 \\ 
        \hline
        $\ket{1}_1$ & 294 & 9227 & 97 & 291 \\
        \hline
        $\ket{2}_1$ & 195 & 121 & 9382 & 271 \\
        \hline
        $\ket{3}_1$ & 132 & 338 & 327 & 9878 \\
        \hline
    \end{tabular}
    \caption{Noiseless subspace Fourier coincidence matrix for $d=4,k=4$.}
    \label{tab:my_label}
\end{table}

\begin{table}[h!]
    \centering
    \begin{tabular}{|c|c|c|} 
        \hline
        Comp. & $\ket{0}_1$ & $\ket{1}_1$ \\ [0.5ex] 
        \hline
        $\ket{0}_1$ & 10562 & 53 \\ 
        \hline
        $\ket{1}_1$ & 47 & 9731 \\
        \hline
    \end{tabular}
    \caption{Noiseless computational coincidence matrix for $d=2$.}
    \label{tab:my_label}
\end{table}

\begin{table}[h!]
    \centering
    \begin{tabular}{|c|c|c|} 
        \hline
        Four. & $\ket{0}_1$ & $\ket{1}_1$ \\ [0.5ex] 
        \hline
        $\ket{0}_1$ & 9867 & 225 \\ 
        \hline
        $\ket{1}_1$ & 216 & 10103 \\
        \hline
    \end{tabular}
    \caption{Noiseless subspace Fourier coincidence matrix for $d=2,k=2$.}
    \label{tab:my_label}
\end{table}

\begin{table}[h!]
    \centering
    \begin{tabular}{|c|c|c|c|c|c|c|c|c|} 
        \hline
        Four. & $\ket{0}_{Bob_1}$ & $\ket{1}_{Bob_1}$ & $\ket{0}_{Bob_2}$ & $\ket{1}_{Bob_2}$ & 
           $\ket{0}_{Bob_3}$ & $\ket{1}_{Bob_3}$ & $\ket{2}_{Bob_3}$ & $\ket{3}_{Bob_3}$ \\ [0.5ex] 
        \hline
        $\ket{0}_{Bob_1}$ & 9569 & 311 & 0 & 0 & 0 & 0 & 0 & 0 \\ 
        \hline
        $\ket{1}_{Bob_1}$ & 290 & 9768 & 0 & 0 & 0 & 0 & 0 & 0 \\
        \hline
        $\ket{0}_{Bob_2}$ & 0 & 0 & 10315 & 318 & 0 & 0 & 0 & 0 \\
        \hline
        $\ket{1}_{Bob_2}$ & 0 & 0 & 271 & 9518 & 0 & 0 & 0 & 0 \\
        \hline
        $\ket{0}_{Bob_3}$ & 0 & 0 & 0 & 0 & 9213 & 301 & 250 & 75 \\  
        \hline 
        $\ket{1}_{Bob_3}$ & 0 & 0 & 0 & 0 & 299 & 9437 & 107 & 261 \\
        \hline
        $\ket{2}_{Bob_3}$ & 0 & 0 & 0 & 0 & 311 & 103 & 9321 & 307 \\
        \hline
        $\ket{3}_{Bob_3}$ & 0 & 0 & 0 & 0 & 108 & 307 & 201 & 9632 \\
        \hline
    \end{tabular}
    \caption{Noiseless subspace Fourier coincidence matrix for "simple QKD network" scenario. The top two $2\times 2$ blocks of this matrix are taken from the $d=8,k=2$ matrix and the bottom $4\times 4$ block from the $d=8,k=4$ scenario.}
    \label{tab:my_label}
\end{table}

\begin{table}[h!]
    \centering
    \begin{tabular}{|c|c|c|c|c|c|c|c|c|} 
        \hline
        Comp. & $\ket{0}$ & $\ket{1}$ & $\ket{2}$ & $\ket{3}$ & 
           $\ket{4}$ & $\ket{5}$ & $\ket{6}$ & $\ket{7}$ \\ [0.5ex] 
        \hline
        $\ket{0}$ & 10372 & 42 & 0 & 0 & 0 & 0 & 0 & 0 \\ 
        \hline
        $\ket{1}$ & 36 & 9641 & 0 & 0 & 0 & 0 & 0 & 0 \\
        \hline
        $\ket{2}$ & 0 & 0 & 10812 & 45 & 0 & 0 & 0 & 0 \\
        \hline
        $\ket{3}$ & 0 & 0 & 41 & 10311 & 0 & 0 & 0 & 0 \\
        \hline
        $\ket{4}$ & 0 & 0 & 0 & 0 & 9715 & 42 & 53 & 41 \\  
        \hline 
        $\ket{5}$ & 0 & 0 & 0 & 0 & 21 & 9621 & 32 & 69 \\
        \hline
        $\ket{6}$ & 0 & 0 & 0 & 0 & 47 & 23 & 10571 & 37 \\
        \hline
        $\ket{7}$ & 0 & 0 & 0 & 0 & 31 & 71 & 41 & 10003 \\
        \hline
    \end{tabular}
    \caption{Noiseless subspace Computational coincidence matrix for "simple QKD network" scenario. The top two $2\times 2$ blocks of this matrix are taken from the $d=8,k=2$ matrix and the bottom $4\times 4$ block from the $d=8,k=4$ scenario.}
    \label{tab:my_label}
\end{table}


\begin{thebibliography}{24}%
\makeatletter
\providecommand \@ifxundefined [1]{%
 \@ifx{#1\undefined}
}%
\providecommand \@ifnum [1]{%
 \ifnum #1\expandafter \@firstoftwo
 \else \expandafter \@secondoftwo
 \fi
}%
\providecommand \@ifx [1]{%
 \ifx #1\expandafter \@firstoftwo
 \else \expandafter \@secondoftwo
 \fi
}%
\providecommand \natexlab [1]{#1}%
\providecommand \enquote  [1]{``#1''}%
\providecommand \bibnamefont  [1]{#1}%
\providecommand \bibfnamefont [1]{#1}%
\providecommand \citenamefont [1]{#1}%
\providecommand \href@noop [0]{\@secondoftwo}%
\providecommand \href [0]{\begingroup \@sanitize@url \@href}%
\providecommand \@href[1]{\@@startlink{#1}\@@href}%
\providecommand \@@href[1]{\endgroup#1\@@endlink}%
\providecommand \@sanitize@url [0]{\catcode `\\12\catcode `\$12\catcode
  `\&12\catcode `\#12\catcode `\^12\catcode `\_12\catcode `\%12\relax}%
\providecommand \@@startlink[1]{}%
\providecommand \@@endlink[0]{}%
\providecommand \url  [0]{\begingroup\@sanitize@url \@url }%
\providecommand \@url [1]{\endgroup\@href {#1}{\urlprefix }}%
\providecommand \urlprefix  [0]{URL }%
\providecommand \Eprint [0]{\href }%
\providecommand \doibase [0]{http://dx.doi.org/}%
\providecommand \selectlanguage [0]{\@gobble}%
\providecommand \bibinfo  [0]{\@secondoftwo}%
\providecommand \bibfield  [0]{\@secondoftwo}%
\providecommand \translation [1]{[#1]}%
\providecommand \BibitemOpen [0]{}%
\providecommand \bibitemStop [0]{}%
\providecommand \bibitemNoStop [0]{.\EOS\space}%
\providecommand \EOS [0]{\spacefactor3000\relax}%
\providecommand \BibitemShut  [1]{\csname bibitem#1\endcsname}%
\let\auto@bib@innerbib\@empty
\bibitem [{\citenamefont {Bennett}\ and\ \citenamefont
  {Brassard}(2014)}]{BENNETT20147}%
  \BibitemOpen
  \bibfield  {author} {\bibinfo {author} {\bibfnamefont {C.~H.}\ \bibnamefont
  {Bennett}}\ and\ \bibinfo {author} {\bibfnamefont {G.}~\bibnamefont
  {Brassard}},\ }\href {\doibase https://doi.org/10.1016/j.tcs.2014.05.025}
  {\bibfield  {journal} {\bibinfo  {journal} {Theoretical Computer Science}\
  }\textbf {\bibinfo {volume} {560}},\ \bibinfo {pages} {7 } (\bibinfo {year}
  {2014})},\ \bibinfo {note} {theoretical Aspects of Quantum Cryptography –
  celebrating 30 years of BB84}\BibitemShut {NoStop}%
\bibitem [{\citenamefont {Ekert}(1991)}]{PhysRevLett.67.661}%
  \BibitemOpen
  \bibfield  {author} {\bibinfo {author} {\bibfnamefont {A.~K.}\ \bibnamefont
  {Ekert}},\ }\href {\doibase 10.1103/PhysRevLett.67.661} {\bibfield  {journal}
  {\bibinfo  {journal} {Phys. Rev. Lett.}\ }\textbf {\bibinfo {volume} {67}},\
  \bibinfo {pages} {661} (\bibinfo {year} {1991})}\BibitemShut {NoStop}%
\bibitem [{\citenamefont {Menezes}\ \emph {et~al.}(2001)\citenamefont
  {Menezes}, \citenamefont {van Oorschot},\ and\ \citenamefont
  {Vanstone}}]{HAC}%
  \BibitemOpen
  \bibfield  {author} {\bibinfo {author} {\bibfnamefont {A.~J.}\ \bibnamefont
  {Menezes}}, \bibinfo {author} {\bibfnamefont {P.~C.}\ \bibnamefont {van
  Oorschot}}, \ and\ \bibinfo {author} {\bibfnamefont {S.~A.}\ \bibnamefont
  {Vanstone}},\ }\href {http://www.cacr.math.uwaterloo.ca/hac/} {\emph
  {\bibinfo {title} {Handbook of Applied Cryptography}}}\ (\bibinfo
  {publisher} {CRC Press},\ \bibinfo {year} {2001})\BibitemShut {NoStop}%
\bibitem [{\citenamefont {Bechmann-Pasquinucci}\ and\ \citenamefont
  {Tittel}(2000)}]{bechmann-pasquinucci2000}%
  \BibitemOpen
  \bibfield  {author} {\bibinfo {author} {\bibfnamefont {H.}~\bibnamefont
  {Bechmann-Pasquinucci}}\ and\ \bibinfo {author} {\bibfnamefont
  {W.}~\bibnamefont {Tittel}},\ }\href {\doibase 10.1103/PhysRevA.61.062308}
  {\bibfield  {journal} {\bibinfo  {journal} {Phys. Rev. A}\ }\textbf {\bibinfo
  {volume} {61}},\ \bibinfo {pages} {062308} (\bibinfo {year}
  {2000})}\BibitemShut {NoStop}%
\bibitem [{\citenamefont {Sheridan}\ and\ \citenamefont
  {Scarani}(2010)}]{PhysRevA.82.030301}%
  \BibitemOpen
  \bibfield  {author} {\bibinfo {author} {\bibfnamefont {L.}~\bibnamefont
  {Sheridan}}\ and\ \bibinfo {author} {\bibfnamefont {V.}~\bibnamefont
  {Scarani}},\ }\href {\doibase 10.1103/PhysRevA.82.030301} {\bibfield
  {journal} {\bibinfo  {journal} {Phys. Rev. A}\ }\textbf {\bibinfo {volume}
  {82}},\ \bibinfo {pages} {030301} (\bibinfo {year} {2010})}\BibitemShut
  {NoStop}%
\bibitem [{\citenamefont {Etcheverry}\ \emph {et~al.}(2013)\citenamefont
  {Etcheverry}, \citenamefont {Ca{\~{n}}as}, \citenamefont {G{\'o}mez},
  \citenamefont {Nogueira}, \citenamefont {Saavedra}, \citenamefont {Xavier},\
  and\ \citenamefont {Lima}}]{Etcheverry2013}%
  \BibitemOpen
  \bibfield  {author} {\bibinfo {author} {\bibfnamefont {S.}~\bibnamefont
  {Etcheverry}}, \bibinfo {author} {\bibfnamefont {G.}~\bibnamefont
  {Ca{\~{n}}as}}, \bibinfo {author} {\bibfnamefont {E.~S.}\ \bibnamefont
  {G{\'o}mez}}, \bibinfo {author} {\bibfnamefont {W.~A.~T.}\ \bibnamefont
  {Nogueira}}, \bibinfo {author} {\bibfnamefont {C.}~\bibnamefont {Saavedra}},
  \bibinfo {author} {\bibfnamefont {G.~B.}\ \bibnamefont {Xavier}}, \ and\
  \bibinfo {author} {\bibfnamefont {G.}~\bibnamefont {Lima}},\ }\href {\doibase
  10.1038/srep02316} {\bibfield  {journal} {\bibinfo  {journal} {Scientific
  Reports}\ }\textbf {\bibinfo {volume} {3}},\ \bibinfo {pages} {2316}
  (\bibinfo {year} {2013})}\BibitemShut {NoStop}%
\bibitem [{\citenamefont {Mirhosseini}\ \emph {et~al.}(2015)\citenamefont
  {Mirhosseini}, \citenamefont {Maga{\~{n}}a-Loaiza}, \citenamefont
  {O'Sullivan}, \citenamefont {Rodenburg}, \citenamefont {Malik}, \citenamefont
  {Lavery}, \citenamefont {Padgett}, \citenamefont {Gauthier},\ and\
  \citenamefont {Boyd}}]{Mirhosseini_2015}%
  \BibitemOpen
  \bibfield  {author} {\bibinfo {author} {\bibfnamefont {M.}~\bibnamefont
  {Mirhosseini}}, \bibinfo {author} {\bibfnamefont {O.~S.}\ \bibnamefont
  {Maga{\~{n}}a-Loaiza}}, \bibinfo {author} {\bibfnamefont {M.~N.}\
  \bibnamefont {O'Sullivan}}, \bibinfo {author} {\bibfnamefont
  {B.}~\bibnamefont {Rodenburg}}, \bibinfo {author} {\bibfnamefont
  {M.}~\bibnamefont {Malik}}, \bibinfo {author} {\bibfnamefont {M.~P.~J.}\
  \bibnamefont {Lavery}}, \bibinfo {author} {\bibfnamefont {M.~J.}\
  \bibnamefont {Padgett}}, \bibinfo {author} {\bibfnamefont {D.~J.}\
  \bibnamefont {Gauthier}}, \ and\ \bibinfo {author} {\bibfnamefont {R.~W.}\
  \bibnamefont {Boyd}},\ }\href {\doibase 10.1088/1367-2630/17/3/033033}
  {\bibfield  {journal} {\bibinfo  {journal} {New Journal of Physics}\ }\textbf
  {\bibinfo {volume} {17}},\ \bibinfo {pages} {033033} (\bibinfo {year}
  {2015})}\BibitemShut {NoStop}%
\bibitem [{\citenamefont {Ca\~nas}\ \emph {et~al.}(2017)\citenamefont
  {Ca\~nas}, \citenamefont {Vera}, \citenamefont {Cari\~ne}, \citenamefont
  {Gonz\'alez}, \citenamefont {Cardenas}, \citenamefont {Connolly},
  \citenamefont {Przysiezna}, \citenamefont {G\'omez}, \citenamefont
  {Figueroa}, \citenamefont {Vallone}, \citenamefont {Villoresi}, \citenamefont
  {da~Silva}, \citenamefont {Xavier},\ and\ \citenamefont
  {Lima}}]{PhysRevA.96.022317}%
  \BibitemOpen
  \bibfield  {author} {\bibinfo {author} {\bibfnamefont {G.}~\bibnamefont
  {Ca\~nas}}, \bibinfo {author} {\bibfnamefont {N.}~\bibnamefont {Vera}},
  \bibinfo {author} {\bibfnamefont {J.}~\bibnamefont {Cari\~ne}}, \bibinfo
  {author} {\bibfnamefont {P.}~\bibnamefont {Gonz\'alez}}, \bibinfo {author}
  {\bibfnamefont {J.}~\bibnamefont {Cardenas}}, \bibinfo {author}
  {\bibfnamefont {P.~W.~R.}\ \bibnamefont {Connolly}}, \bibinfo {author}
  {\bibfnamefont {A.}~\bibnamefont {Przysiezna}}, \bibinfo {author}
  {\bibfnamefont {E.~S.}\ \bibnamefont {G\'omez}}, \bibinfo {author}
  {\bibfnamefont {M.}~\bibnamefont {Figueroa}}, \bibinfo {author}
  {\bibfnamefont {G.}~\bibnamefont {Vallone}}, \bibinfo {author} {\bibfnamefont
  {P.}~\bibnamefont {Villoresi}}, \bibinfo {author} {\bibfnamefont {T.~F.}\
  \bibnamefont {da~Silva}}, \bibinfo {author} {\bibfnamefont {G.~B.}\
  \bibnamefont {Xavier}}, \ and\ \bibinfo {author} {\bibfnamefont
  {G.}~\bibnamefont {Lima}},\ }\href {\doibase 10.1103/PhysRevA.96.022317}
  {\bibfield  {journal} {\bibinfo  {journal} {Phys. Rev. A}\ }\textbf {\bibinfo
  {volume} {96}},\ \bibinfo {pages} {022317} (\bibinfo {year}
  {2017})}\BibitemShut {NoStop}%
\bibitem [{\citenamefont {Islam}\ \emph {et~al.}(2017)\citenamefont {Islam},
  \citenamefont {Lim}, \citenamefont {Cahall}, \citenamefont {Kim},\ and\
  \citenamefont {Gauthier}}]{islam2017provably}%
  \BibitemOpen
  \bibfield  {author} {\bibinfo {author} {\bibfnamefont {N.~T.}\ \bibnamefont
  {Islam}}, \bibinfo {author} {\bibfnamefont {C.~C.~W.}\ \bibnamefont {Lim}},
  \bibinfo {author} {\bibfnamefont {C.}~\bibnamefont {Cahall}}, \bibinfo
  {author} {\bibfnamefont {J.}~\bibnamefont {Kim}}, \ and\ \bibinfo {author}
  {\bibfnamefont {D.~J.}\ \bibnamefont {Gauthier}},\ }\href
  {https://advances.sciencemag.org/content/3/11/e1701491} {\bibfield  {journal}
  {\bibinfo  {journal} {Science advances}\ }\textbf {\bibinfo {volume} {3}},\
  \bibinfo {pages} {e1701491} (\bibinfo {year} {2017})}\BibitemShut {NoStop}%
\bibitem [{\citenamefont {Sit}\ \emph {et~al.}(2017)\citenamefont {Sit},
  \citenamefont {Bouchard}, \citenamefont {Fickler}, \citenamefont
  {Gagnon-Bischoff}, \citenamefont {Larocque}, \citenamefont {Heshami},
  \citenamefont {Elser}, \citenamefont {Peuntinger}, \citenamefont
  {G\"{u}nthner}, \citenamefont {Heim}, \citenamefont {Marquardt},
  \citenamefont {Leuchs}, \citenamefont {Boyd},\ and\ \citenamefont
  {Karimi}}]{Sit:17}%
  \BibitemOpen
  \bibfield  {author} {\bibinfo {author} {\bibfnamefont {A.}~\bibnamefont
  {Sit}}, \bibinfo {author} {\bibfnamefont {F.}~\bibnamefont {Bouchard}},
  \bibinfo {author} {\bibfnamefont {R.}~\bibnamefont {Fickler}}, \bibinfo
  {author} {\bibfnamefont {J.}~\bibnamefont {Gagnon-Bischoff}}, \bibinfo
  {author} {\bibfnamefont {H.}~\bibnamefont {Larocque}}, \bibinfo {author}
  {\bibfnamefont {K.}~\bibnamefont {Heshami}}, \bibinfo {author} {\bibfnamefont
  {D.}~\bibnamefont {Elser}}, \bibinfo {author} {\bibfnamefont
  {C.}~\bibnamefont {Peuntinger}}, \bibinfo {author} {\bibfnamefont
  {K.}~\bibnamefont {G\"{u}nthner}}, \bibinfo {author} {\bibfnamefont
  {B.}~\bibnamefont {Heim}}, \bibinfo {author} {\bibfnamefont {C.}~\bibnamefont
  {Marquardt}}, \bibinfo {author} {\bibfnamefont {G.}~\bibnamefont {Leuchs}},
  \bibinfo {author} {\bibfnamefont {R.~W.}\ \bibnamefont {Boyd}}, \ and\
  \bibinfo {author} {\bibfnamefont {E.}~\bibnamefont {Karimi}},\ }\href
  {\doibase 10.1364/OPTICA.4.001006} {\bibfield  {journal} {\bibinfo  {journal}
  {Optica}\ }\textbf {\bibinfo {volume} {4}},\ \bibinfo {pages} {1006}
  (\bibinfo {year} {2017})}\BibitemShut {NoStop}%
\bibitem [{\citenamefont {Ding}\ \emph {et~al.}(2017)\citenamefont {Ding},
  \citenamefont {Bacco}, \citenamefont {Dalgaard}, \citenamefont {Cai},
  \citenamefont {Zhou}, \citenamefont {Rottwitt},\ and\ \citenamefont
  {Oxenl{\o}we}}]{Ding2017}%
  \BibitemOpen
  \bibfield  {author} {\bibinfo {author} {\bibfnamefont {Y.}~\bibnamefont
  {Ding}}, \bibinfo {author} {\bibfnamefont {D.}~\bibnamefont {Bacco}},
  \bibinfo {author} {\bibfnamefont {K.}~\bibnamefont {Dalgaard}}, \bibinfo
  {author} {\bibfnamefont {X.}~\bibnamefont {Cai}}, \bibinfo {author}
  {\bibfnamefont {X.}~\bibnamefont {Zhou}}, \bibinfo {author} {\bibfnamefont
  {K.}~\bibnamefont {Rottwitt}}, \ and\ \bibinfo {author} {\bibfnamefont
  {L.~K.}\ \bibnamefont {Oxenl{\o}we}},\ }\href {\doibase
  10.1038/s41534-017-0026-2} {\bibfield  {journal} {\bibinfo  {journal} {npj
  Quantum Information}\ }\textbf {\bibinfo {volume} {3}},\ \bibinfo {pages}
  {25} (\bibinfo {year} {2017})}\BibitemShut {NoStop}%
\bibitem [{\citenamefont {Cozzolino}\ \emph {et~al.}(2019)\citenamefont
  {Cozzolino}, \citenamefont {Bacco}, \citenamefont {Da~Lio}, \citenamefont
  {Ingerslev}, \citenamefont {Ding}, \citenamefont {Dalgaard}, \citenamefont
  {Kristensen}, \citenamefont {Galili}, \citenamefont {Rottwitt}, \citenamefont
  {Ramachandran},\ and\ \citenamefont
  {Oxenl\o{}we}}]{PhysRevApplied.11.064058}%
  \BibitemOpen
  \bibfield  {author} {\bibinfo {author} {\bibfnamefont {D.}~\bibnamefont
  {Cozzolino}}, \bibinfo {author} {\bibfnamefont {D.}~\bibnamefont {Bacco}},
  \bibinfo {author} {\bibfnamefont {B.}~\bibnamefont {Da~Lio}}, \bibinfo
  {author} {\bibfnamefont {K.}~\bibnamefont {Ingerslev}}, \bibinfo {author}
  {\bibfnamefont {Y.}~\bibnamefont {Ding}}, \bibinfo {author} {\bibfnamefont
  {K.}~\bibnamefont {Dalgaard}}, \bibinfo {author} {\bibfnamefont
  {P.}~\bibnamefont {Kristensen}}, \bibinfo {author} {\bibfnamefont
  {M.}~\bibnamefont {Galili}}, \bibinfo {author} {\bibfnamefont
  {K.}~\bibnamefont {Rottwitt}}, \bibinfo {author} {\bibfnamefont
  {S.}~\bibnamefont {Ramachandran}}, \ and\ \bibinfo {author} {\bibfnamefont
  {L.~K.}\ \bibnamefont {Oxenl\o{}we}},\ }\href {\doibase
  10.1103/PhysRevApplied.11.064058} {\bibfield  {journal} {\bibinfo  {journal}
  {Phys. Rev. Applied}\ }\textbf {\bibinfo {volume} {11}},\ \bibinfo {pages}
  {064058} (\bibinfo {year} {2019})}\BibitemShut {NoStop}%
\bibitem [{\citenamefont {Islam}\ \emph {et~al.}(2019)\citenamefont {Islam},
  \citenamefont {Lim}, \citenamefont {Cahall}, \citenamefont {Qi},
  \citenamefont {Kim},\ and\ \citenamefont {Gauthier}}]{Islam_2019}%
  \BibitemOpen
  \bibfield  {author} {\bibinfo {author} {\bibfnamefont {N.~T.}\ \bibnamefont
  {Islam}}, \bibinfo {author} {\bibfnamefont {C.~C.~W.}\ \bibnamefont {Lim}},
  \bibinfo {author} {\bibfnamefont {C.}~\bibnamefont {Cahall}}, \bibinfo
  {author} {\bibfnamefont {B.}~\bibnamefont {Qi}}, \bibinfo {author}
  {\bibfnamefont {J.}~\bibnamefont {Kim}}, \ and\ \bibinfo {author}
  {\bibfnamefont {D.~J.}\ \bibnamefont {Gauthier}},\ }\href {\doibase
  10.1088/2058-9565/ab21a4} {\bibfield  {journal} {\bibinfo  {journal} {Quantum
  Science and Technology}\ }\textbf {\bibinfo {volume} {4}},\ \bibinfo {pages}
  {035008} (\bibinfo {year} {2019})}\BibitemShut {NoStop}%
\bibitem [{\citenamefont {Vagniluca}\ \emph {et~al.}(2020)\citenamefont
  {Vagniluca}, \citenamefont {Da~Lio}, \citenamefont {Rusca}, \citenamefont
  {Cozzolino}, \citenamefont {Ding}, \citenamefont {Zbinden}, \citenamefont
  {Zavatta}, \citenamefont {Oxenl\o{}we},\ and\ \citenamefont
  {Bacco}}]{PhysRevApplied.14.014051}%
  \BibitemOpen
  \bibfield  {author} {\bibinfo {author} {\bibfnamefont {I.}~\bibnamefont
  {Vagniluca}}, \bibinfo {author} {\bibfnamefont {B.}~\bibnamefont {Da~Lio}},
  \bibinfo {author} {\bibfnamefont {D.}~\bibnamefont {Rusca}}, \bibinfo
  {author} {\bibfnamefont {D.}~\bibnamefont {Cozzolino}}, \bibinfo {author}
  {\bibfnamefont {Y.}~\bibnamefont {Ding}}, \bibinfo {author} {\bibfnamefont
  {H.}~\bibnamefont {Zbinden}}, \bibinfo {author} {\bibfnamefont
  {A.}~\bibnamefont {Zavatta}}, \bibinfo {author} {\bibfnamefont {L.~K.}\
  \bibnamefont {Oxenl\o{}we}}, \ and\ \bibinfo {author} {\bibfnamefont
  {D.}~\bibnamefont {Bacco}},\ }\href {\doibase
  10.1103/PhysRevApplied.14.014051} {\bibfield  {journal} {\bibinfo  {journal}
  {Phys. Rev. Applied}\ }\textbf {\bibinfo {volume} {14}},\ \bibinfo {pages}
  {014051} (\bibinfo {year} {2020})}\BibitemShut {NoStop}%
\bibitem [{\citenamefont {Zhong}\ \emph {et~al.}(2015)\citenamefont {Zhong},
  \citenamefont {Zhou}, \citenamefont {Horansky}, \citenamefont {Lee},
  \citenamefont {Verma}, \citenamefont {Lita}, \citenamefont {Restelli},
  \citenamefont {Bienfang}, \citenamefont {Mirin}, \citenamefont {Gerrits},
  \citenamefont {Nam}, \citenamefont {Marsili}, \citenamefont {Shaw},
  \citenamefont {Zhang}, \citenamefont {Wang}, \citenamefont {Englund},
  \citenamefont {Wornell}, \citenamefont {Shapiro},\ and\ \citenamefont
  {Wong}}]{Zhong_2015}%
  \BibitemOpen
  \bibfield  {author} {\bibinfo {author} {\bibfnamefont {T.}~\bibnamefont
  {Zhong}}, \bibinfo {author} {\bibfnamefont {H.}~\bibnamefont {Zhou}},
  \bibinfo {author} {\bibfnamefont {R.~D.}\ \bibnamefont {Horansky}}, \bibinfo
  {author} {\bibfnamefont {C.}~\bibnamefont {Lee}}, \bibinfo {author}
  {\bibfnamefont {V.~B.}\ \bibnamefont {Verma}}, \bibinfo {author}
  {\bibfnamefont {A.~E.}\ \bibnamefont {Lita}}, \bibinfo {author}
  {\bibfnamefont {A.}~\bibnamefont {Restelli}}, \bibinfo {author}
  {\bibfnamefont {J.~C.}\ \bibnamefont {Bienfang}}, \bibinfo {author}
  {\bibfnamefont {R.~P.}\ \bibnamefont {Mirin}}, \bibinfo {author}
  {\bibfnamefont {T.}~\bibnamefont {Gerrits}}, \bibinfo {author} {\bibfnamefont
  {S.~W.}\ \bibnamefont {Nam}}, \bibinfo {author} {\bibfnamefont
  {F.}~\bibnamefont {Marsili}}, \bibinfo {author} {\bibfnamefont {M.~D.}\
  \bibnamefont {Shaw}}, \bibinfo {author} {\bibfnamefont {Z.}~\bibnamefont
  {Zhang}}, \bibinfo {author} {\bibfnamefont {L.}~\bibnamefont {Wang}},
  \bibinfo {author} {\bibfnamefont {D.}~\bibnamefont {Englund}}, \bibinfo
  {author} {\bibfnamefont {G.~W.}\ \bibnamefont {Wornell}}, \bibinfo {author}
  {\bibfnamefont {J.~H.}\ \bibnamefont {Shapiro}}, \ and\ \bibinfo {author}
  {\bibfnamefont {F.~N.~C.}\ \bibnamefont {Wong}},\ }\href {\doibase
  10.1088/1367-2630/17/2/022002} {\bibfield  {journal} {\bibinfo  {journal}
  {New Journal of Physics}\ }\textbf {\bibinfo {volume} {17}},\ \bibinfo
  {pages} {022002} (\bibinfo {year} {2015})}\BibitemShut {NoStop}%
\bibitem [{\citenamefont {Ecker}\ \emph {et~al.}(2019)\citenamefont {Ecker},
  \citenamefont {Bouchard}, \citenamefont {Bulla}, \citenamefont {Brandt},
  \citenamefont {Kohout}, \citenamefont {Steinlechner}, \citenamefont
  {Fickler}, \citenamefont {Malik}, \citenamefont {Guryanova}, \citenamefont
  {Ursin},\ and\ \citenamefont {Huber}}]{PhysRevX.9.041042}%
  \BibitemOpen
  \bibfield  {author} {\bibinfo {author} {\bibfnamefont {S.}~\bibnamefont
  {Ecker}}, \bibinfo {author} {\bibfnamefont {F.}~\bibnamefont {Bouchard}},
  \bibinfo {author} {\bibfnamefont {L.}~\bibnamefont {Bulla}}, \bibinfo
  {author} {\bibfnamefont {F.}~\bibnamefont {Brandt}}, \bibinfo {author}
  {\bibfnamefont {O.}~\bibnamefont {Kohout}}, \bibinfo {author} {\bibfnamefont
  {F.}~\bibnamefont {Steinlechner}}, \bibinfo {author} {\bibfnamefont
  {R.}~\bibnamefont {Fickler}}, \bibinfo {author} {\bibfnamefont
  {M.}~\bibnamefont {Malik}}, \bibinfo {author} {\bibfnamefont
  {Y.}~\bibnamefont {Guryanova}}, \bibinfo {author} {\bibfnamefont
  {R.}~\bibnamefont {Ursin}}, \ and\ \bibinfo {author} {\bibfnamefont
  {M.}~\bibnamefont {Huber}},\ }\href {\doibase 10.1103/PhysRevX.9.041042}
  {\bibfield  {journal} {\bibinfo  {journal} {Phys. Rev. X}\ }\textbf {\bibinfo
  {volume} {9}},\ \bibinfo {pages} {041042} (\bibinfo {year}
  {2019})}\BibitemShut {NoStop}%
\bibitem [{\citenamefont {Doda}\ \emph {et~al.}(2021)\citenamefont {Doda},
  \citenamefont {Huber}, \citenamefont {Murta}, \citenamefont {Pivoluska},
  \citenamefont {Plesch},\ and\ \citenamefont {Vlachou}}]{doda2020quantum}%
  \BibitemOpen
  \bibfield  {author} {\bibinfo {author} {\bibfnamefont {M.}~\bibnamefont
  {Doda}}, \bibinfo {author} {\bibfnamefont {M.}~\bibnamefont {Huber}},
  \bibinfo {author} {\bibfnamefont {G.}~\bibnamefont {Murta}}, \bibinfo
  {author} {\bibfnamefont {M.}~\bibnamefont {Pivoluska}}, \bibinfo {author}
  {\bibfnamefont {M.}~\bibnamefont {Plesch}}, \ and\ \bibinfo {author}
  {\bibfnamefont {C.}~\bibnamefont {Vlachou}},\ }\href {\doibase
  10.1103/PhysRevApplied.15.034003} {\bibfield  {journal} {\bibinfo  {journal}
  {Phys. Rev. Applied}\ }\textbf {\bibinfo {volume} {15}},\ \bibinfo {pages}
  {034003} (\bibinfo {year} {2021})}\BibitemShut {NoStop}%
\bibitem [{\citenamefont {Hu}\ \emph {et~al.}(2021)\citenamefont {Hu},
  \citenamefont {Zhang}, \citenamefont {Guo}, \citenamefont {Wang},
  \citenamefont {Xing}, \citenamefont {Huang}, \citenamefont {Liu},
  \citenamefont {Huang}, \citenamefont {Li}, \citenamefont {Guo}, \citenamefont
  {Gao}, \citenamefont {Pivoluska},\ and\ \citenamefont
  {Huber}}]{PhysRevLett.127.110505}%
  \BibitemOpen
  \bibfield  {author} {\bibinfo {author} {\bibfnamefont {X.-M.}\ \bibnamefont
  {Hu}}, \bibinfo {author} {\bibfnamefont {C.}~\bibnamefont {Zhang}}, \bibinfo
  {author} {\bibfnamefont {Y.}~\bibnamefont {Guo}}, \bibinfo {author}
  {\bibfnamefont {F.-X.}\ \bibnamefont {Wang}}, \bibinfo {author}
  {\bibfnamefont {W.-B.}\ \bibnamefont {Xing}}, \bibinfo {author}
  {\bibfnamefont {C.-X.}\ \bibnamefont {Huang}}, \bibinfo {author}
  {\bibfnamefont {B.-H.}\ \bibnamefont {Liu}}, \bibinfo {author} {\bibfnamefont
  {Y.-F.}\ \bibnamefont {Huang}}, \bibinfo {author} {\bibfnamefont {C.-F.}\
  \bibnamefont {Li}}, \bibinfo {author} {\bibfnamefont {G.-C.}\ \bibnamefont
  {Guo}}, \bibinfo {author} {\bibfnamefont {X.}~\bibnamefont {Gao}}, \bibinfo
  {author} {\bibfnamefont {M.}~\bibnamefont {Pivoluska}}, \ and\ \bibinfo
  {author} {\bibfnamefont {M.}~\bibnamefont {Huber}},\ }\href {\doibase
  10.1103/PhysRevLett.127.110505} {\bibfield  {journal} {\bibinfo  {journal}
  {Phys. Rev. Lett.}\ }\textbf {\bibinfo {volume} {127}},\ \bibinfo {pages}
  {110505} (\bibinfo {year} {2021})}\BibitemShut {NoStop}%
\bibitem [{\citenamefont {Ou}(2007)}]{ou2007multi}%
  \BibitemOpen
  \bibfield  {author} {\bibinfo {author} {\bibfnamefont {Z.-Y.~J.}\
  \bibnamefont {Ou}},\ }\href {https://www.springer.com/gp/book/9780387255323}
  {\emph {\bibinfo {title} {Multi-photon quantum interference}}},\
  Vol.~\bibinfo {volume} {43}\ (\bibinfo  {publisher} {Springer},\ \bibinfo
  {year} {2007})\BibitemShut {NoStop}%
\bibitem [{\citenamefont {Schneeloch}\ \emph {et~al.}(2019)\citenamefont
  {Schneeloch}, \citenamefont {Knarr}, \citenamefont {Bogorin}, \citenamefont
  {Levangie}, \citenamefont {Tison}, \citenamefont {Frank}, \citenamefont
  {Howland}, \citenamefont {Fanto},\ and\ \citenamefont
  {Alsing}}]{schneeloch2019introduction}%
  \BibitemOpen
  \bibfield  {author} {\bibinfo {author} {\bibfnamefont {J.}~\bibnamefont
  {Schneeloch}}, \bibinfo {author} {\bibfnamefont {S.~H.}\ \bibnamefont
  {Knarr}}, \bibinfo {author} {\bibfnamefont {D.~F.}\ \bibnamefont {Bogorin}},
  \bibinfo {author} {\bibfnamefont {M.~L.}\ \bibnamefont {Levangie}}, \bibinfo
  {author} {\bibfnamefont {C.~C.}\ \bibnamefont {Tison}}, \bibinfo {author}
  {\bibfnamefont {R.}~\bibnamefont {Frank}}, \bibinfo {author} {\bibfnamefont
  {G.~A.}\ \bibnamefont {Howland}}, \bibinfo {author} {\bibfnamefont {M.~L.}\
  \bibnamefont {Fanto}}, \ and\ \bibinfo {author} {\bibfnamefont {P.~M.}\
  \bibnamefont {Alsing}},\ }\href {\doibase 10.1088/2040-8986/ab05a8}
  {\bibfield  {journal} {\bibinfo  {journal} {Journal of Optics}\ }\textbf
  {\bibinfo {volume} {21}},\ \bibinfo {pages} {043501} (\bibinfo {year}
  {2019})}\BibitemShut {NoStop}%
\bibitem [{\citenamefont {Ma}\ \emph {et~al.}(2007)\citenamefont {Ma},
  \citenamefont {Fung},\ and\ \citenamefont {Lo}}]{PhysRevA.76.012307}%
  \BibitemOpen
  \bibfield  {author} {\bibinfo {author} {\bibfnamefont {X.}~\bibnamefont
  {Ma}}, \bibinfo {author} {\bibfnamefont {C.-H.~F.}\ \bibnamefont {Fung}}, \
  and\ \bibinfo {author} {\bibfnamefont {H.-K.}\ \bibnamefont {Lo}},\ }\href
  {\doibase 10.1103/PhysRevA.76.012307} {\bibfield  {journal} {\bibinfo
  {journal} {Phys. Rev. A}\ }\textbf {\bibinfo {volume} {76}},\ \bibinfo
  {pages} {012307} (\bibinfo {year} {2007})}\BibitemShut {NoStop}%
\bibitem [{\citenamefont {Neumann}\ \emph {et~al.}(2021)\citenamefont
  {Neumann}, \citenamefont {Scheidl}, \citenamefont {Selimovic}, \citenamefont
  {Pivoluska}, \citenamefont {Liu}, \citenamefont {Bohmann},\ and\
  \citenamefont {Ursin}}]{PhysRevA.104.022406}%
  \BibitemOpen
  \bibfield  {author} {\bibinfo {author} {\bibfnamefont {S.~P.}\ \bibnamefont
  {Neumann}}, \bibinfo {author} {\bibfnamefont {T.}~\bibnamefont {Scheidl}},
  \bibinfo {author} {\bibfnamefont {M.}~\bibnamefont {Selimovic}}, \bibinfo
  {author} {\bibfnamefont {M.}~\bibnamefont {Pivoluska}}, \bibinfo {author}
  {\bibfnamefont {B.}~\bibnamefont {Liu}}, \bibinfo {author} {\bibfnamefont
  {M.}~\bibnamefont {Bohmann}}, \ and\ \bibinfo {author} {\bibfnamefont
  {R.}~\bibnamefont {Ursin}},\ }\href {\doibase 10.1103/PhysRevA.104.022406}
  {\bibfield  {journal} {\bibinfo  {journal} {Phys. Rev. A}\ }\textbf {\bibinfo
  {volume} {104}},\ \bibinfo {pages} {022406} (\bibinfo {year}
  {2021})}\BibitemShut {NoStop}%
\bibitem [{\citenamefont {Wengerowsky}\ \emph {et~al.}(2018)\citenamefont
  {Wengerowsky}, \citenamefont {Joshi}, \citenamefont {Steinlechner},
  \citenamefont {H{\"u}bel},\ and\ \citenamefont {Ursin}}]{Wengerowsky2018}%
  \BibitemOpen
  \bibfield  {author} {\bibinfo {author} {\bibfnamefont {S.}~\bibnamefont
  {Wengerowsky}}, \bibinfo {author} {\bibfnamefont {S.~K.}\ \bibnamefont
  {Joshi}}, \bibinfo {author} {\bibfnamefont {F.}~\bibnamefont {Steinlechner}},
  \bibinfo {author} {\bibfnamefont {H.}~\bibnamefont {H{\"u}bel}}, \ and\
  \bibinfo {author} {\bibfnamefont {R.}~\bibnamefont {Ursin}},\ }\href
  {\doibase 10.1038/s41586-018-0766-y} {\bibfield  {journal} {\bibinfo
  {journal} {Nature}\ }\textbf {\bibinfo {volume} {564}},\ \bibinfo {pages}
  {225} (\bibinfo {year} {2018})}\BibitemShut {NoStop}%
\bibitem [{\citenamefont {{Koduru Joshi}}\ \emph {et~al.}(2020)\citenamefont
  {{Koduru Joshi}}, \citenamefont {{Aktas}}, \citenamefont {{Wengerowsky}},
  \citenamefont {{Lon{\v{c}}ari{\'c}}}, \citenamefont {{Neumann}},
  \citenamefont {{Liu}}, \citenamefont {{Scheidl}}, \citenamefont {{Curr{\'a}s
  Lorenzo}}, \citenamefont {{Samec}}, \citenamefont {{Kling}}, \citenamefont
  {{Qiu}}, \citenamefont {{Razavi}}, \citenamefont {{Stip{\v{c}}evi{\'c}}},
  \citenamefont {{Rarity}},\ and\ \citenamefont {{Ursin}}}]{Joshi2020}%
  \BibitemOpen
  \bibfield  {author} {\bibinfo {author} {\bibfnamefont {S.}~\bibnamefont
  {{Koduru Joshi}}}, \bibinfo {author} {\bibfnamefont {D.}~\bibnamefont
  {{Aktas}}}, \bibinfo {author} {\bibfnamefont {S.}~\bibnamefont
  {{Wengerowsky}}}, \bibinfo {author} {\bibfnamefont {M.}~\bibnamefont
  {{Lon{\v{c}}ari{\'c}}}}, \bibinfo {author} {\bibfnamefont {S.~P.}\
  \bibnamefont {{Neumann}}}, \bibinfo {author} {\bibfnamefont {B.}~\bibnamefont
  {{Liu}}}, \bibinfo {author} {\bibfnamefont {T.}~\bibnamefont {{Scheidl}}},
  \bibinfo {author} {\bibfnamefont {G.}~\bibnamefont {{Curr{\'a}s Lorenzo}}},
  \bibinfo {author} {\bibfnamefont {{\v{Z}}.}~\bibnamefont {{Samec}}}, \bibinfo
  {author} {\bibfnamefont {L.}~\bibnamefont {{Kling}}}, \bibinfo {author}
  {\bibfnamefont {A.}~\bibnamefont {{Qiu}}}, \bibinfo {author} {\bibfnamefont
  {M.}~\bibnamefont {{Razavi}}}, \bibinfo {author} {\bibfnamefont
  {M.}~\bibnamefont {{Stip{\v{c}}evi{\'c}}}}, \bibinfo {author} {\bibfnamefont
  {J.~G.}\ \bibnamefont {{Rarity}}}, \ and\ \bibinfo {author} {\bibfnamefont
  {R.}~\bibnamefont {{Ursin}}},\ }\href {\doibase 10.1126/sciadv.aba0959}
  {\bibfield  {journal} {\bibinfo  {journal} {Science Advances}\ }\textbf
  {\bibinfo {volume} {6}},\ \bibinfo {pages} {eaba0959} (\bibinfo {year}
  {2020})},\ \Eprint
  {http://arxiv.org/abs/https://www.science.org/doi/pdf/10.1126/sciadv.aba0959}
  {https://www.science.org/doi/pdf/10.1126/sciadv.aba0959} \BibitemShut
  {NoStop}%
\end{thebibliography}%
\end{document}